\documentclass[onecolumn,helvetica]{aastex631}
\def\kms{\hbox{km$\;$s$^{-1}$}}

\DeclareRobustCommand{\ion}[2]{\textup{#1\,\textsc{\lowercase{#2}}}}
\newcommand{\IRIS}{IRIS}
\newcommand{\SDO}{SDO}
\newcommand{\angstrom}{\textup{\AA}}
\newcommand{\lambdat}{$\lambda$$t$}
\newcommand{\lambday}{$\lambda$$y$}
\newcommand{\SiIV}{{\ion{Si}{IV}}}

\newcommand{\MgII}{{\ion{Mg}{II}}}

\usepackage{mwe}

\usepackage{hyperref}
\usepackage{gensymb}
\usepackage{soul}
\hypersetup{linkcolor=blue,citecolor=blue,filecolor=cyan,urlcolor=blue}
\accepted{\today}

\submitjournal{ApJL}

\graphicspath{ {./Figures/} }
\begin{document}

\title{On the million-degree signature of spicules}

\correspondingauthor{Souvik Bose}
\email{bose@lmsal.com}

\author[0000-0002-2180-1013]{Souvik Bose}
\affiliation{Lockheed Martin Solar \& Astrophysics Laboratory, Palo Alto, CA 94304, USA}
\affiliation{SETI Institute, 339 Bernardo Ave, Mountain View, CA 94043, USA}
\affiliation{Institute of Theoretical Astrophysics, University of Oslo, PO Box 1029, Blindern 0315, Oslo, Norway}
\affiliation{Rosseland Centre for Solar Physics, University of Oslo, PO Box 1029, Blindern 0315, Oslo, Norway}

\author[0000-0003-0585-7030]{Jayant Joshi}
\affiliation{Indian Institute of Astrophysics, Koramangala--2nd Block, Bangalore 560034, India}

\author[0000-0002-0405-0668]{Paola Testa}
\affiliation{Harvard-Smithsonian Center for Astrophysics, 60 Garden St, Cambridge, MA 02193, USA}
\author[0000-0002-8370-952X]{Bart De Pontieu}
\affiliation{Lockheed Martin Solar \& Astrophysics Laboratory, Palo Alto, CA 94304, USA}
\affiliation{Institute of Theoretical Astrophysics, University of Oslo, PO Box 1029, Blindern 0315, Oslo, Norway}
\affiliation{Rosseland Centre for Solar Physics, University of Oslo, PO Box 1029, Blindern 0315, Oslo, Norway}

\begin{abstract}

Spicules have often been proposed as substantial contributors toward the mass and energy balance of the solar corona. While their transition region (TR) counterpart has unequivocally been established over the past decade, the observations concerning the coronal contribution of spicules have often been contested. This is mainly attributed to the lack of adequate coordinated observations, their small spatial scales, highly dynamic nature, and complex multi-thermal evolution, which are often observed at the limit of our current observational facilities. Therefore, it remains unclear how much heating occurs in association with spicules to coronal temperatures. In this study, we use coordinated high-resolution observations of the solar chromosphere, TR, and corona of a quiet Sun region and a coronal hole with the Interface Region Imaging Spectrograph (IRIS) and the Atmospheric Imaging Assembly (AIA) to investigate the (lower) coronal ($\sim$1MK) emission associated with spicules. We perform differential emission measure (DEM) analysis on the AIA passbands using basis pursuit and a newly developed technique based on Tikhonov regularization to probe the thermal structure of the spicular environment at coronal temperatures. We find that the EM maps at 1~MK reveal the presence of ubiquitous, small-scale jets with a clear spatio-temporal coherence with the spicules observed in the IRIS/TR passband. Detailed space-time analysis of the chromospheric, TR, and EM maps show unambiguous evidence of rapidly outward propagating spicules with strong emission (2—3 times higher than the background) at 1~MK. Our findings are consistent with previously reported MHD simulations that show heating to coronal temperatures associated with spicules.   

\end{abstract}

\keywords{Solar Physics (1476) --- The Sun (1693) --- Solar atmosphere (1477) --- Solar Corona (1483) --- Solar spicules (1525) --- Solar chromosphere (1479) --- Solar coronal heating (1989)}

\section{Introduction} \label{sec:intro}

Spicules are thin, dynamic, thread-like features that appear ubiquitously on the surface of the Sun. They are one of the most abundantly observed features in the chromosphere, and their origin and role have long been a subject of debate \citep{1972ARA&A..10...73B,2000_sterling_review,2019_hinode_review,2019ARA&A..57..189C}. Being ubiquitous, the mechanisms that drive spicules have held promise as contributing to coronal heating events \citep{1982ApJ...255..743A,2011Sci...331...55D}, and the chromospheric mass-flux that these events propel to coronal heights is estimated to be two orders of magnitude higher than required to balance the mass-loss due to solar wind \citep{1983ApJ...267..825W}. It is estimated that at any given moment, the Sun's surface hosts at least a million spicules \citep[][and possibly significantly higher based on high-resolution observations]{1972ARA&A..10...73B} in active regions or quiet Sun, \citep[][]{2010ApJ...719..469J} rapidly propagating outward. 

Historically, spicules have mainly been observed in the chromospheric and transition region (TR) passband \citep{1972ARA&A..10...73B,1989SoPh..123...41D}, but due to the lack of adequate high-resolution observations, a coronal counterpart was missing until high-resolution extreme ultraviolet (EUV) observations from space became available in the past decade. As a result, the possibility of energizing the solar corona through spicules was dismissed as unlikely in many of these early studies. The discovery of the short-lived and more dynamic ($\sim$50$-$100~km/s) off-limb ``type-II'' spicules by \cite{2007PASJ...59S.655D} sparked renewed interest in their role of outer atmospheric heating since many of these spicules appeared to ``fade'' from the Hinode \ion{Ca}{ii}~H passband unlike their \st{classical} type-I counterpart. Such fading suggested a scenario where the opacity of \ion{Ca}{ii}~H type-II spicules rapidly dropped during the spicule lifetime, possibly because of a combination of their dynamic evolution or even heating, to higher temperatures. Coordinated observations between Hinode and Atmospheric Imaging Assembly \citep[AIA,][]{2012SoPh..275...17L} channels onboard NASA's Solar Dynamics Observatory \citep[SDO,][]{2012SoPh..275....3P}, and later with the Interface Region Imaging Spectrograph \citep[IRIS,][]{2014SoPh..289.2733D} revealed that significant heating occurs in at least a subset of threads in type-II spicules along their whole length to TR temperatures \citep[$\sim$80,000~K,][]{2011Sci...331...55D,2014ApJ...792L..15P}. The on-disk counterparts of type-II spicules, termed rapid blue/red-shifted excursions \citep[RBEs and RREs,][]{2009ApJ...705..272R}, were also associated with heating to at least TR \citep[termed as network jets,][]{2014Sci...346A.315T,2015ApJ...799L...3R} and possibly even coronal temperatures \citep{2011Sci...331...55D,2016ApJ...820..124H,2017ApJ...845L..18D}.

Despite substantial advancements in our understanding of the impact of spicules in the TR, their role in mass loading and heating the corona has remained a subject of significant debate both from an observational and a theoretical point of view. Several studies challenged their importance for the coronal mass and energy balance and argued that type-II spicules either play no role in coronal emission/heating \citep{2011A&A...532L...1M} or their role is likely \textit{not} a dominant one \citep{2013ApJ...779....1T}. This continued controversy is, in part, because the relatively poor resolution (compared to the size of the spicules) of existing coronal instruments has rendered it challenging to assess their impact on the coronal energy balance. Moreover, the presence of cooler (TR, $<$0.5~MK) ions in the optically thin lines in the AIA passbands \citep{2010A&A...521A..21O,2011ApJ...743...23M,2018LRSP...15....5D} renders an additional challenge because the observed emission could well be attributed to these ``cooler'' ions instead of $\geq$1~MK coronal emission. 
Furthermore, simplifying numerical assumptions \citep{2012JGRA..11712102K,2022_shanwlee} on the nature of the spicular plasma and the single-field-line approach of modeling spicules underestimates the complexity of the spicular environment, as evidenced by the complex processes involved in the coronal heating associated with spicules in multi-dimensional radiative MHD simulations \citep{2017Sci...356.1269M}.

The focus of this paper is not on the contribution of spicules towards coronal heating but rather we take a step back and attempt to investigate their $\sim$1~MK signature unambiguously, by targeting a quiet Sun (QS) and a coronal hole (CH) region, using coordinated IRIS and SDO/AIA observations. By narrowing our target to the above regions, we are most likely studying the $\sim$1~MK signature associated with type-II spicules since they are the more abundantly found in QS and CHs \citep{2012ApJ...759...18P}. Therefore, unless otherwise mentioned, we refer to type-II spicules generally as spicules in this paper. We exploit the high-resolution observations from IRIS to track the TR counterpart of (chromospheric) spicules and investigate their impact on the associated coronal structures, which can be observed in the form of propagating coronal disturbances \citep[PCDs,][]{2010_bdp_scott_pcds,2015_pcds_samanta,2016ApJ...829L..18Bryans,2023ApJ...944..171B}. Differential emission measure (DEM) analysis using two independent approaches is performed to study the thermal structure of the spicular plasma/PCDs within a temperature range centered around $\sim$1~MK.

\section{Observations and data analysis} \label{sec:obs_and_data}

\begin{figure*}
\gridline{\fig{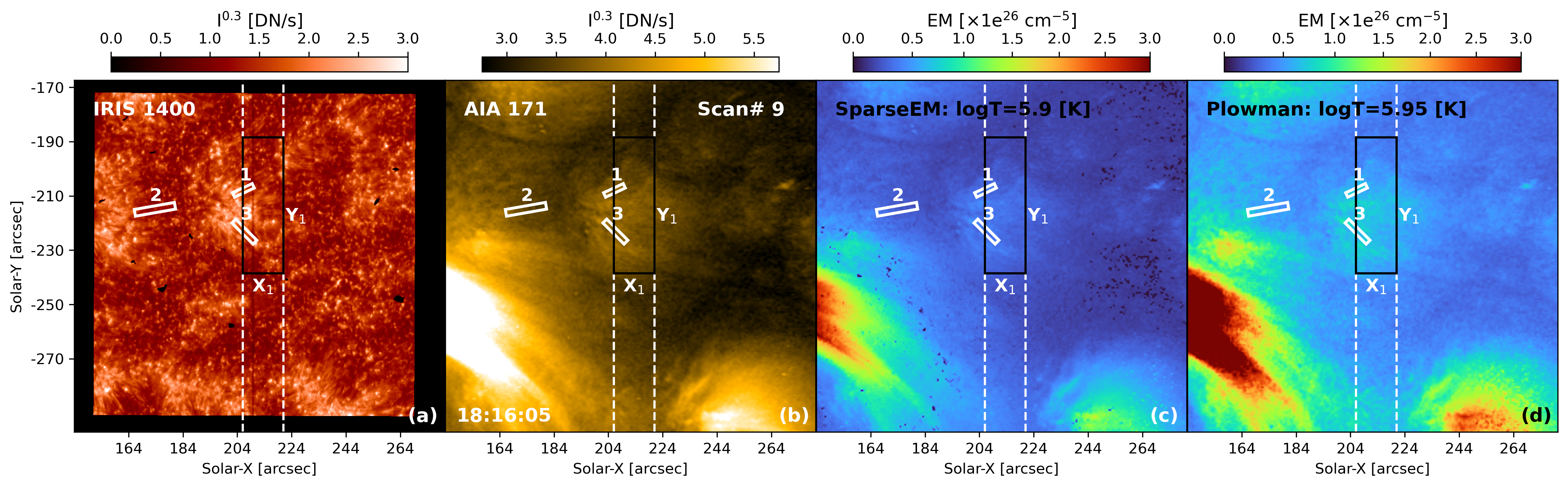}{0.95\textwidth}{}}
\gridline{\fig{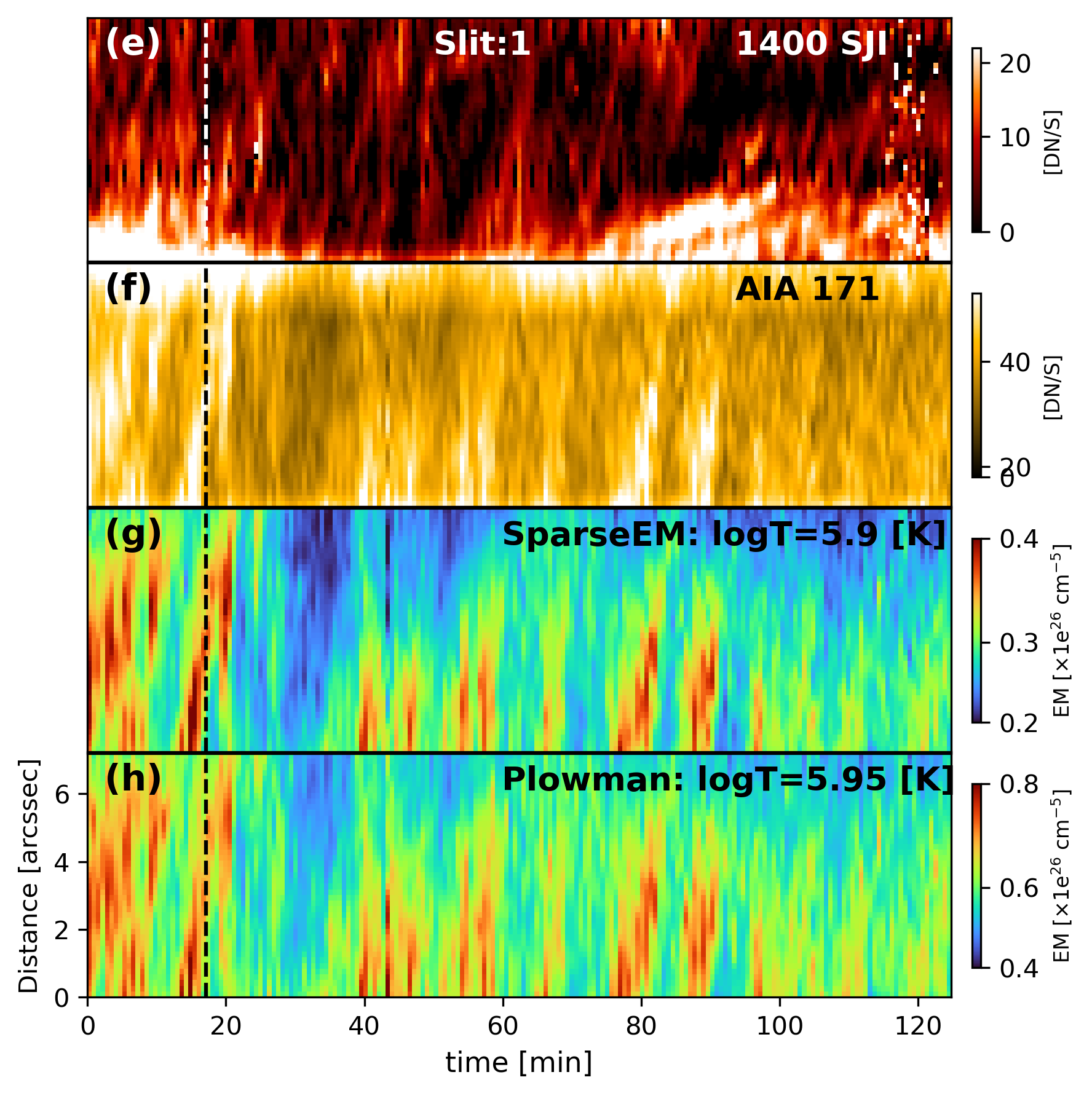}{0.33\textwidth}{}
          \fig{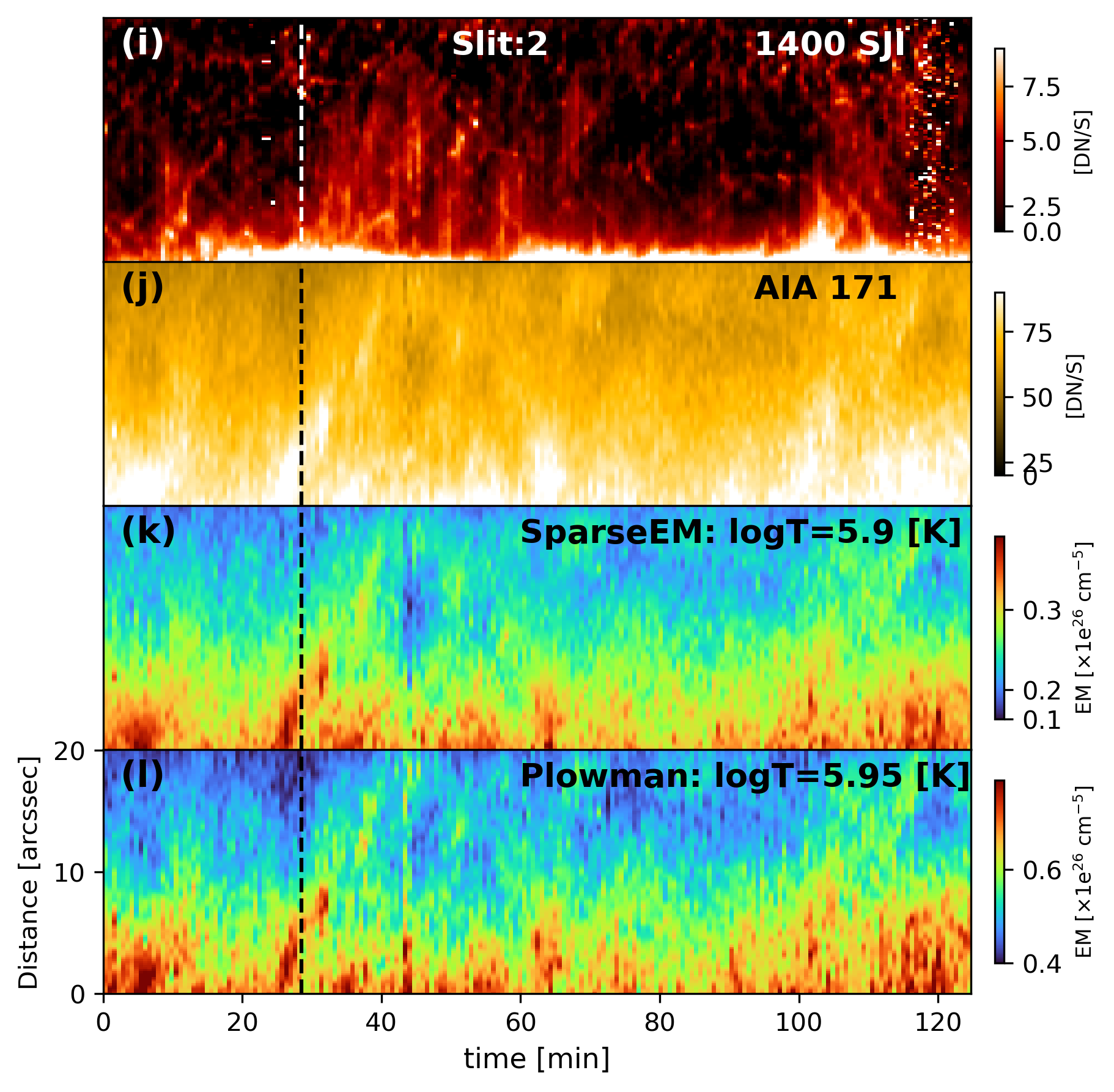}{0.33\textwidth}{}
          \fig{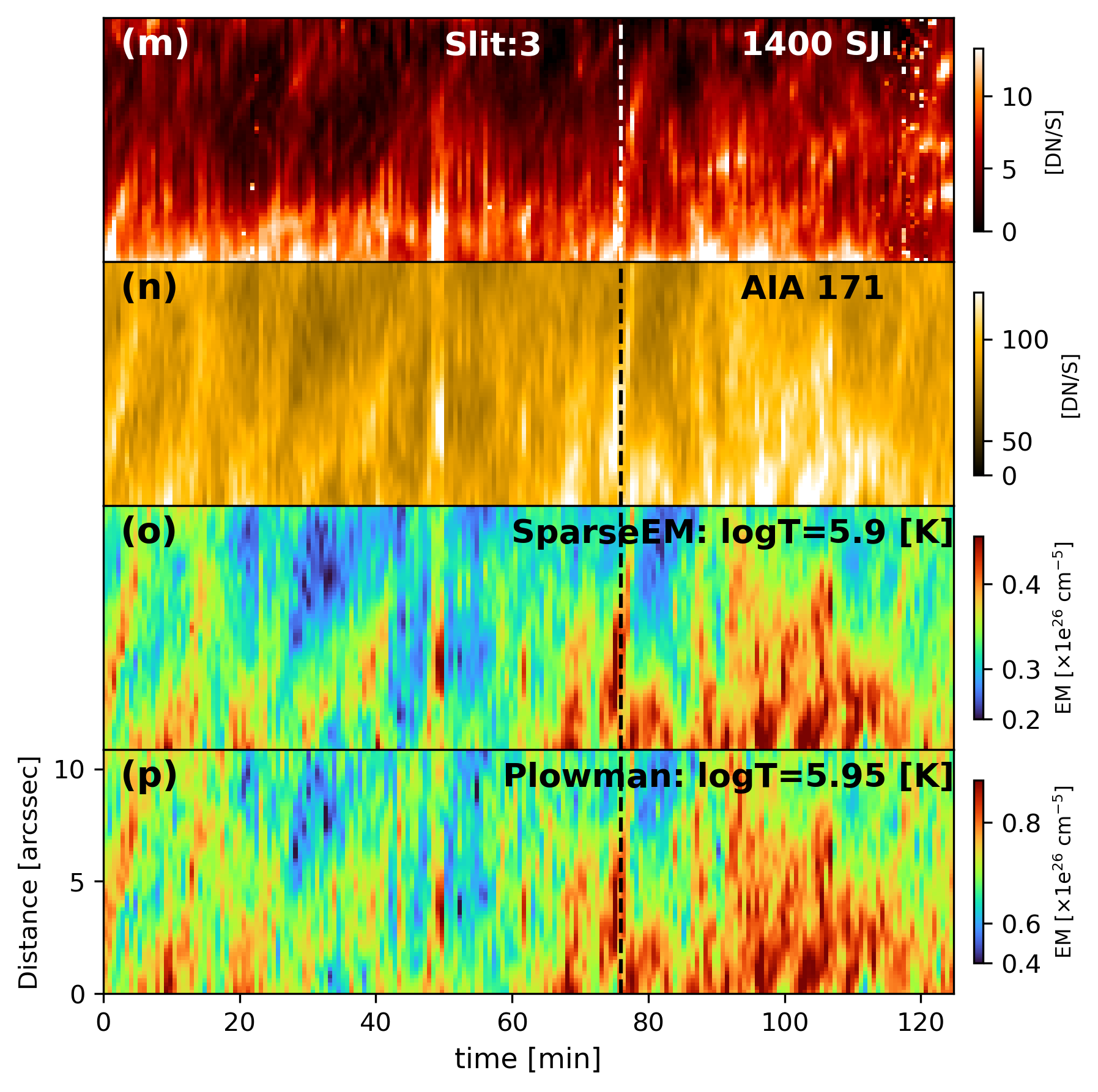}{0.33\textwidth}{}}
    \centering
    \caption{Overview of the coordinated IRIS--SDO/AIA quiet-Sun dataset~1 and the derived space-time maps. \textit{Top row, panel~(a)}: IRIS \SiIV\ 1400~\AA\ SJ image, \textit{panel~(b)}: corresponding SDO/AIA~171~\AA\ image, \textit{panel~(c)}: EM map at log~T[K]=5.9 computed from the sparse-matrix method \citep{2015ApJ...807..143C}, \textit{panel~(d)}: EM map at log~T[K]=5.95 obtained from the Tikhonov regularization method \citep{2020ApJ...905...17P}. The white dashed vertical lines in the above panels indicate the FOV covered by the IRIS rasters and the rectangular region drawn in black indicates the region, which is further analyzed in Figure~\ref{fig:synth_rast_spectra_ds1}. \textit{Panel~(e)}: \SiIV\ 1400~SJI space-time map showing the propagation of the spicular plasma and their associated emission in the form of linear ridge-like structures derived using artificial slit ``1'', \textit{panel~(f)}: corresponding SDO/AIA~171~\AA\ space-time map, \textit{panels~(g) and (h)}: space-time maps corresponding to EM derived using the methods described in \cite{2015ApJ...807..143C} and \cite{2020ApJ...905...17P}, respectively. \textit{Panels~(i--l) and (m--p)} are the same as panels~(e--h) but for artificial slits ``2'' and ``3''. The white/black dashed vertical lines \textit{panels~(e)--(p)} correspond to the time step at which these maps are shown. Animations of this figure are available \href{https://www.dropbox.com/scl/fo/lo1tweijruwfvn9j9vztl/AD3IFStjtgUv5gDGh5Pae7U?rlkey=cst82l1yi8hdb5p90vnvcf3m0&st=8rw2deoq&dl=0}{online} which shows the evolution of multiple network jets in the dataset over 2 hrs and 6 min in particular along the three artificial slits.}
    \label{fig:gridline1}
\end{figure*}

\begin{figure*}
\gridline{\fig{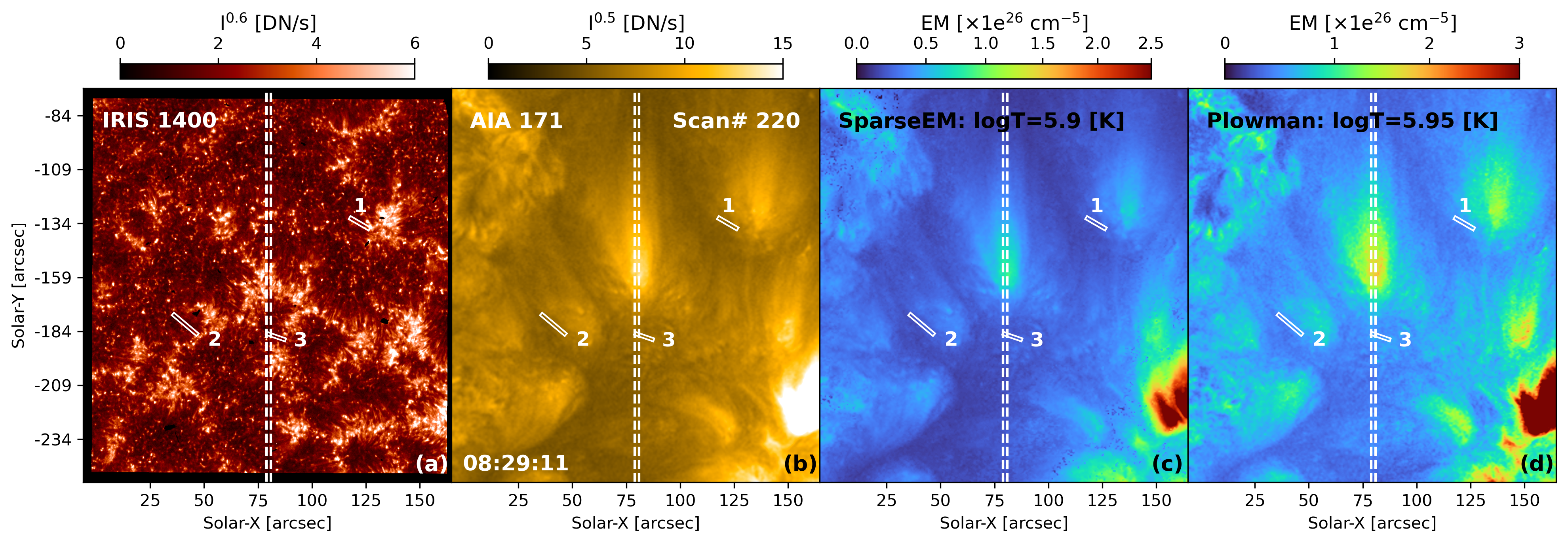}{0.95\textwidth}{}}
\gridline{\fig{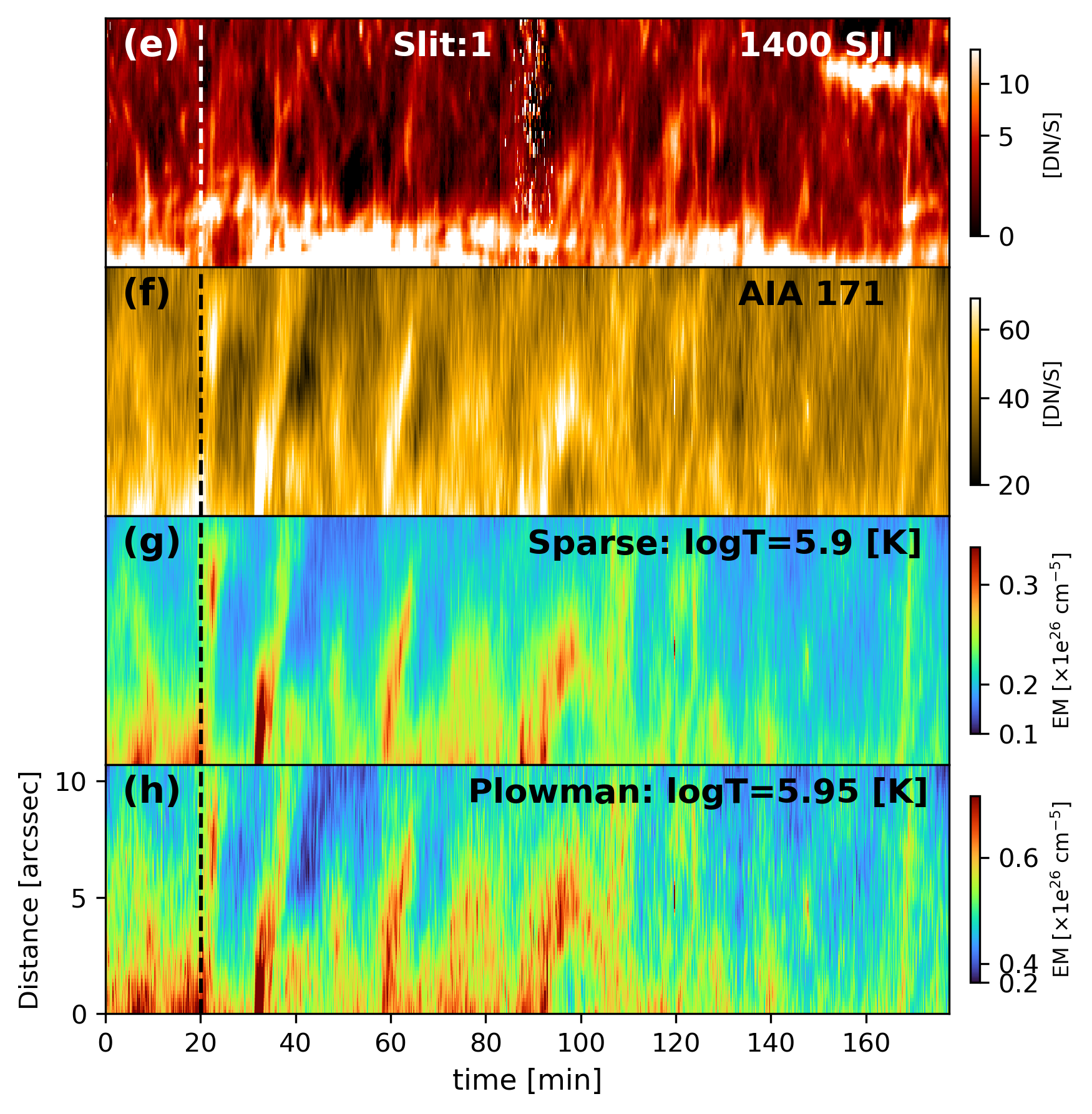}{0.33\textwidth}{}
          \fig{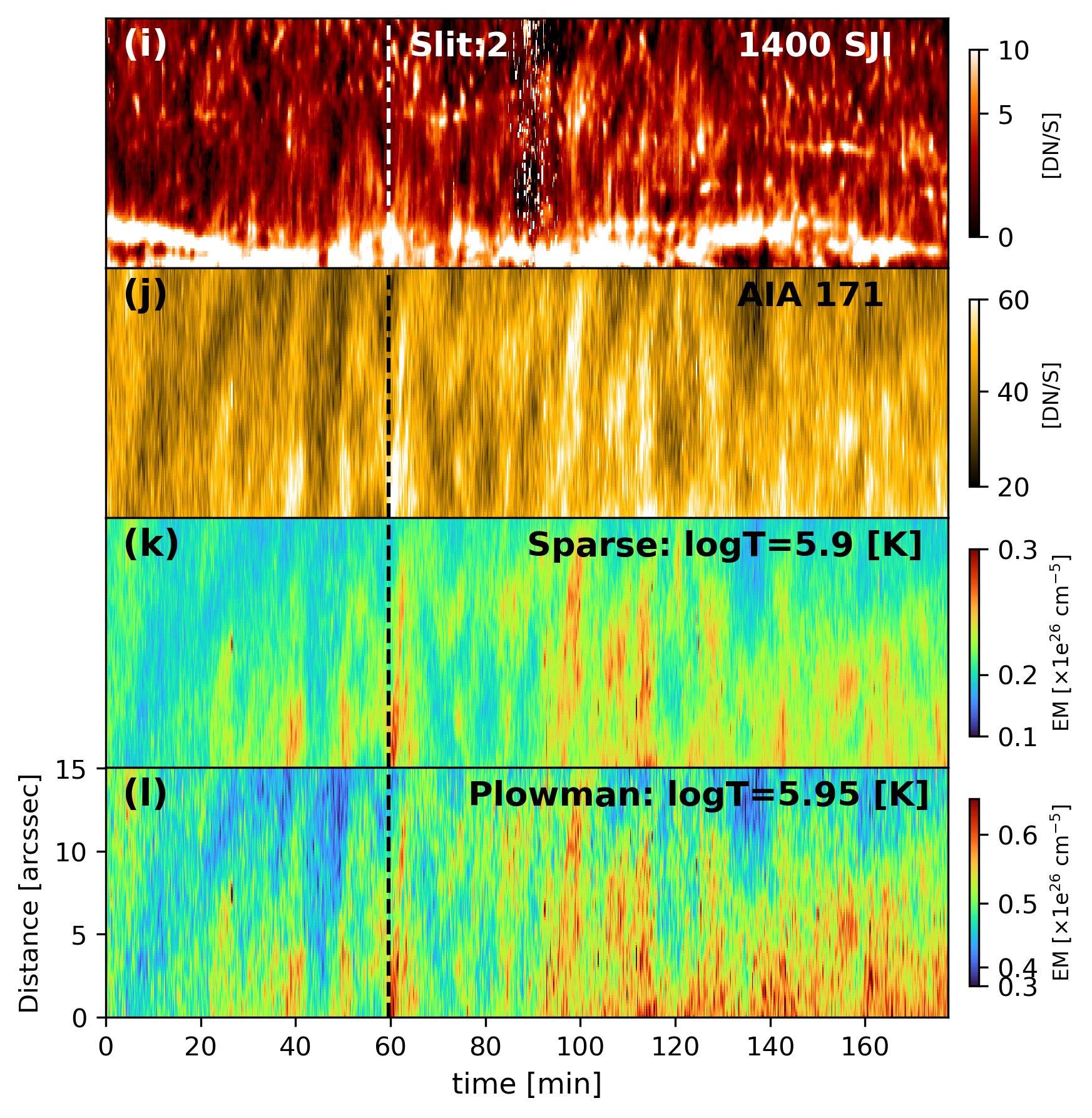}{0.33\textwidth}{}
          \fig{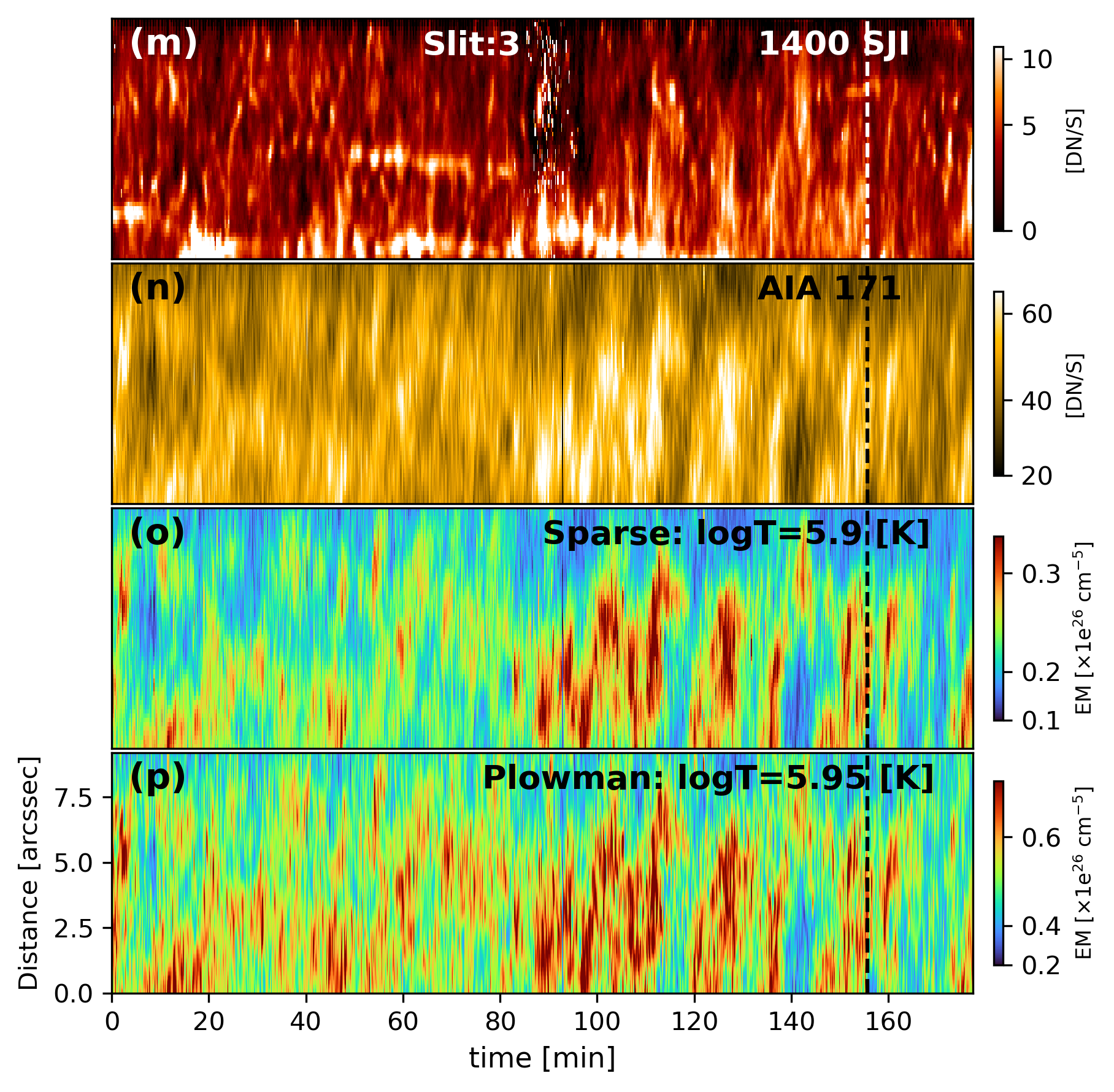}{0.345\textwidth}{}}
    \centering
    \caption{Overview of the coordinated IRIS--SDO/AIA coronal hole dataset~2 and the derived space-time maps. The format of this figure is the same as Figure~\ref{fig:gridline1} above. The FOV bounded by the IRIS raster is further analyzed in Figure~\ref{fig:synth_rast_spectra_ds2}. Animations of this figure are also available \href{https://www.dropbox.com/scl/fo/j8fpv8ywizz6z15zawakr/AHqikfyJgln3KQg7AY0TCUY?rlkey=wfw1of70kkm4yauqpkzskz4gv&st=wutjucey&dl=0}{online} showing the solar evolution and the dynamics of the network jets for 3 hrs in the dataset in the same format as Fig.~\ref{fig:gridline1}  }.
    \label{fig:gridline2}
\end{figure*}

We use two coordinated IRIS-SDO/AIA observations, from 24 September 2014, targeting QS (henceforth dataset~1) and CH (henceforth dataset 2) regions. IRIS ran in a large sparse 16-step raster mode (\verb|OBS id: 3823009186|) targeting a QS region (Fig.~\ref{fig:gridline1}~a) centered around solar ($X$,$Y$)=(211\arcsec,-238\arcsec) with $\mu$=0.94 in dataset~1 ($\mu$ being the cosine of the heliocentric angle). The duration of the dataset was 2~hrs and 6~min starting at 18:09~UTC. Though IRIS provides spectra and simultaneous slit-jaw images (SJIs) in several spectral windows \citep[see][for details]{2014SoPh..289.2733D}, we concentrated on the chromospheric \MgII~2796~\AA\ and the TR dominated \SiIV~1400~\AA\ SJIs in this study that had a field-of-view (FOV) spanning 120\arcsec$\times$120\arcsec, a cadence of 38~s and a pixel scale of 0\farcs16, along with the rasters that had a cadence of 150~s (with a step cadence of 9.5~s, a step size of 1\arcsec\ and an exposure time of 8~s per slit position). The FOV covered by the rasters was 15\arcsec$\times$120\arcsec~in the direction perpendicular and parallel to the slit direction as indicated in Fig.~\ref{fig:gridline1}(a). Dataset~2 was recorded in a very large dense 4-step raster mode (\verb|OBS id: 3820257466|) targeting a CH region (Fig.~\ref{fig:gridline2}~a) around solar ($X$,$Y$)=(78\arcsec,-167\arcsec) with a roll angle of 90\degree\ in the counter-clockwise direction. The value of $\mu$ was 0.98. The observed duration was $\approx$~3~hrs starting at 07:49~UTC. The \MgII~2796~\AA\ and \SiIV~1400~\AA\ SJIs were recorded at a cadence of 11~s and with a FOV of 167\arcsec$\times$174\arcsec. The rasters were acquired at a cadence of 21~s (with a step cadence of 5.4~s and 4~s exposure), with a step size of 0\farcs35 in the direction perpendicular to the slit covering a FOV of 1\arcsec$\times$174\arcsec.

For each of the two \IRIS\ datasets, we downloaded the co-temporal SDO/AIA observations to investigate the extreme ultraviolet (EUV) response associated with spicules. These datasets were prepped, co-aligned, and normalized using the standard \verb|aiapy| \citep{Barnes2020} routines. Additionally, the AIA images corresponding to dataset~2 were rolled by 90\degree\ to have the same orientation as IRIS. The AIA data were further cropped, expanded (to IRIS SJI pixel scale), and spatially and temporally aligned to the respective IRIS SJIs by cross-correlating the (AIA) 1600~\AA\ and (IRIS) \MgII~2796~\AA\ channels. Panel~(b) of Figs.~\ref{fig:gridline1} and \ref{fig:gridline2} show the co-aligned AIA~171~\AA\ images. We used the AIA data at the original pixel scale of 0\farcs6 for the DEM analysis. We defined 31 temperature bins from log~T[K]=5.5 to log~T[K]=7 in steps of 0.05 in log space for the Tikhonov regularization method by \cite{2020ApJ...905...17P}, and 21 temperature bins between log~T[K]=5.7 and log~T[K]=7.7 in steps of 0.1 (also in log space) for the sparse matrix-based method by \cite{2015ApJ...807..143C}. The resulting EM maps, obtained from the two codes, were later co-aligned to IRIS SJIs using the same parameters (shifts, crop, interpolation, etc.) as the AIA images. They are shown in panels~(c) and (d) of Figs.~\ref{fig:gridline1} and \ref{fig:gridline2}. The co-aligned cubes were extensively visualized and analyzed with CRISPEX \citep{2012ApJ...750...22V}, which is an IDL widget-based tool to visualize multi-dimensional data.

The propagation of spicules across the chromospheric, TR, and coronal channels was visualized by performing a space-time analysis along multiple artificial slits in both SJIs and raster maps as shown in Figs.~\ref{fig:gridline1}, \ref{fig:gridline2} and \ref{fig:synth_rast_spectra_ds1}. The slits were chosen after visually inspecting the animation of the co-aligned datasets around the network regions where spicules are most abundant. Each slit is 10 IRIS pixels (1\farcs6) wide and has variable lengths ranging from 7--20\arcsec. For the space-time analysis on the raster, synthetic AIA, and EM rasters were made that were spatially and temporally co-aligned with the respective IRIS rasters. This was done to ensure consistency between the spectrograph observations, which need some time to ``build'' the FOV while the AIA and EM images are instantaneous. 

The IRIS \MgII~k and \SiIV~1402.77~\AA\ spectra were fitted with double and single Gaussian functions, respectively, to extract the corresponding intensities, Doppler shifts, and \MgII~k$_{2}$ peak separation. We use the double-Gaussian fitting technique similar to the one employed by \citet{2015ApJ...811..127S}. Spicules on the solar disk are known to appear in absorption in the chromospheric \MgII\ spectra and have large line-of-sight (LOS) velocity gradients along with enhanced opacities and mass flows \citep{2019A&A...631L...5B}, which causes an enhancement in the k$_{2}$ peak separation \citep{2013ApJ...778..143P,2016ApJ...829L..18Bryans}. This makes visualizing spicules in k$_{2}$ peak separation maps easier than the intensity images. In the TR, spicules appear in emission in the form of network jets \citep{2014Sci...346A.315T} with enhanced peak emission and non-thermal line broadening \citep{2015ApJ...799L...3R}. In this paper, we focus on the \SiIV\ peak and the \MgII~k$_{3}$ intensity and Doppler shift, along with the k$_{2}$ peak separation.

\section{Results} \label{sec:results}

\begin{figure}[!htb]
    \centering
    \includegraphics[width=0.95\textwidth,height=20.5cm]{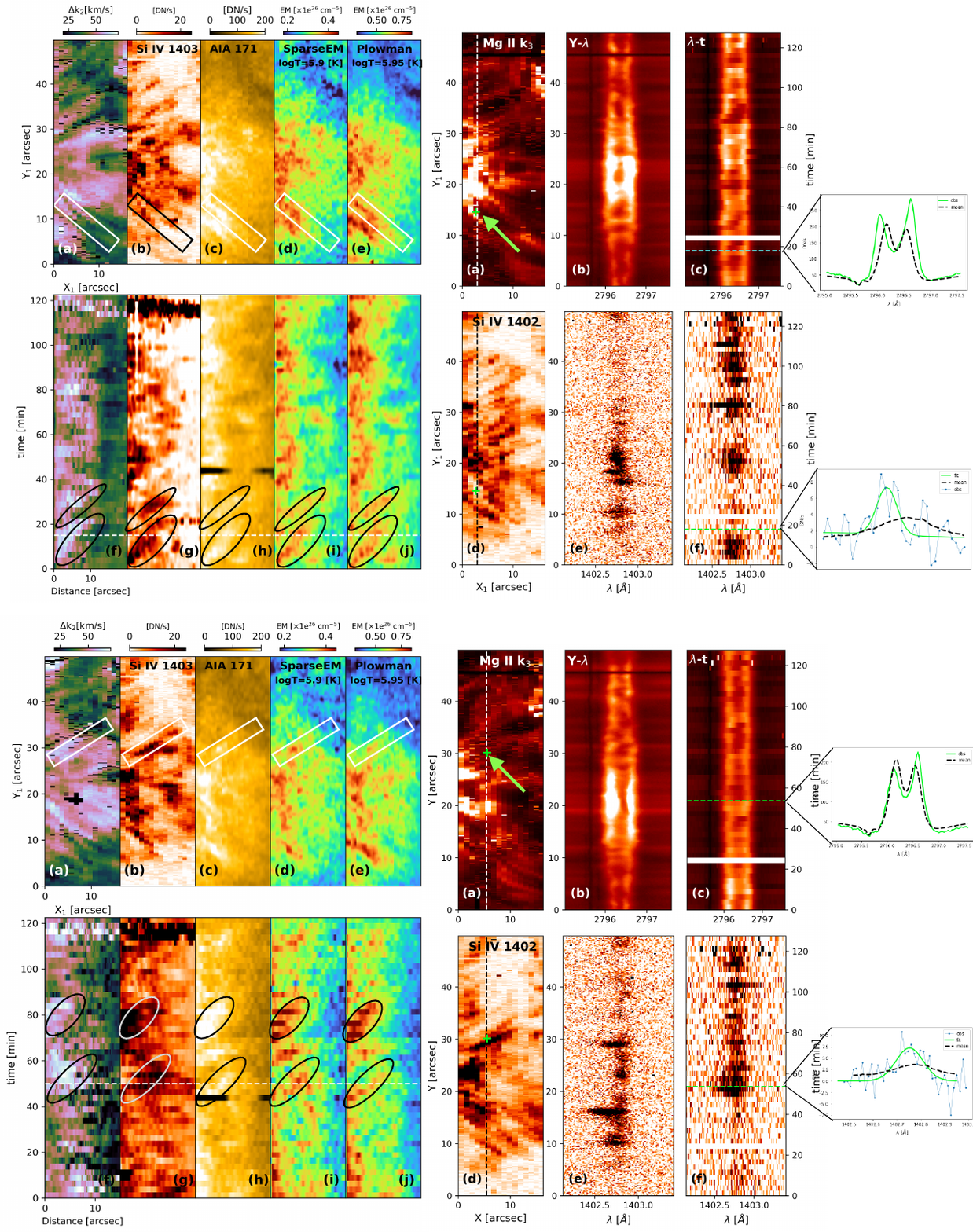}
    \caption{Propagation of spicules in the chromosphere, TR, and lower corona ($\sim$1MK) as observed from a portion of the raster map (black rectangular region in Figure~\ref{fig:gridline1}) and their corresponding \MgII~k and \SiIV\ spectra. \textit{Left column, panel~(a):} \MgII~k$_{2}$ peak-separation map, \textit{panel~(b):} peak intensity map derived from \SiIV\ 1402.77~\AA\ spectra, \textit{panel~(c):} synthetic SDO/AIA~171 rasters, \textit{panel~(d):} synthetic sparse EM rasters, \textit{panel~(e):} synthetic Tikhonov EM rasters, corresponding to the IRIS rasters. The rectangular boxes in \textit{(a)--(e)} show the FOV from which the corresponding space-time maps shown in \textit{panels~(f)--(j)} are derived. The ovals shown in the space-time maps exemplify the chromospheric-TR-lower coronal connection associated with some spicules. \textit{Right column, panel~(a)}: intensity in \MgII~k$_{3}$ derived from bi-Gaussian fit, \textit{panel~(d)}: peak intensity of \SiIV\ 1402.77~\AA\ spectra derived from single-Gaussian fit, \textit{panel~(b)}: spectral-space (\lambday) diagram of \MgII, \textit{panel~(e)}: \lambday\ diagram of \SiIV, obtained along with vertical slice denoted by dashed lines in \textit{panels~(a) and (d)}; \textit{panel~(c)}: corresponding \MgII\ spectral-time (\lambdat) diagram, \textit{panel~(f)}: corresponding \SiIV\ \lambdat\ diagram of the location indicated by the green arrow in \textit{panels~(a)}. The inset figures show the \MgII\ and \SiIV\ spectra at a couple of instances in time. Animations of this figure are also available \href{https://www.dropbox.com/scl/fo/g7fha48leyp11mc7w8spn/AOi1nPxzi1D3UqMqfFndSjo?rlkey=l2ohndg5rim9yz2d7hbl77lrf&st=8n4r29fe&dl=0}{online}.}
    \label{fig:synth_rast_spectra_ds1}
\end{figure}

The animation associated with \SiIV~1400 and AIA~171 channels in the top rows of Figs.~\ref{fig:gridline1} and \ref{fig:gridline2} show that the network regions at the footpoints of coronal loops are replete with spicule-like features. This scenario is consistent with several studies conducted in the past such as, \cite{2016ApJ...829L..18Bryans,2017ApJ...845L..18D,2023ApJ...944..171B}, where the coronal loops show significant complexities and are traditionally associated with propagating coronal disturbances (PCDs). The spicule-like features, also termed as network jets in the TR \citep{2014Sci...346A.315T,2015ApJ...799L...3R}, can have apparent speeds in the range 40--200~\kms \citep{2016SoPh..291.1129N} and rapidly propagate outwards. This is often followed by a downflowing (returning) phase \citep{1983ApJ...267..825W,2021A&A...654A..51B,2021A&A...647A.147B} where the spicules are seen to retract after reaching their maximum extent.
% , although this may not always be possible to observe \citep[e.g. refer to][]{2020ApJ...889...95M,2021A&A...654A..51B}.

The reconstructed EM maps at $\sim$log~T[K]=$5.9$ (panels~c and d of Figs.~\ref{fig:gridline1} and \ref{fig:gridline2}) and their animation show features akin to AIA~171 observations. This is not surprising since the temperature bins of the EM maps shown in the figures are close to the peak temperature response  \citep[log~T(K)$\approx$5.8,][]{2012SoPh..275...41B}{} of the 171 channel. However, unlike the AIA observations, which are often contaminated with ions formed at cooler (TR) temperatures \citep{2010A&A...521A..21O,2011ApJ...743...23M,2018LRSP...15....5D}, these maps show the amount of emission over the whole FOV (integrated along the line-of-sight) within a narrow temperature bin centered at $\sim$log~T[K]=$5.9$ and are therefore better constrained than individual images. However, despite their widespread usage, EM inversions are not fully reliable at temperatures below log~T[K]$\approx$5.6 \citep[refer to the discussions in][]{2015ApJ...807..143C} due to the potential, but ambiguous, low TR contributions to the AIA channels. Using two independent techniques, based on isothermal approximation \citep[e.g.][]{2007_Cirtain_EM_loci} and filter-ratio diagnostics \citep[e.g.,][]{2011_Narukage_filter_ratio,2020_Testa_filer_ratio}, we find that the minimum temperature associated with the emission of the spicular plasma at $\sim$log~T[K]=$5.9$ bin cannot be lower than log~T[K]=5.7 ($\approx$500~kK). We have discussed this in detail in Appendix~\ref{sec:appendix_lower_T}.

\begin{figure}[ht!]
    \centering
    \includegraphics[width=\textwidth]{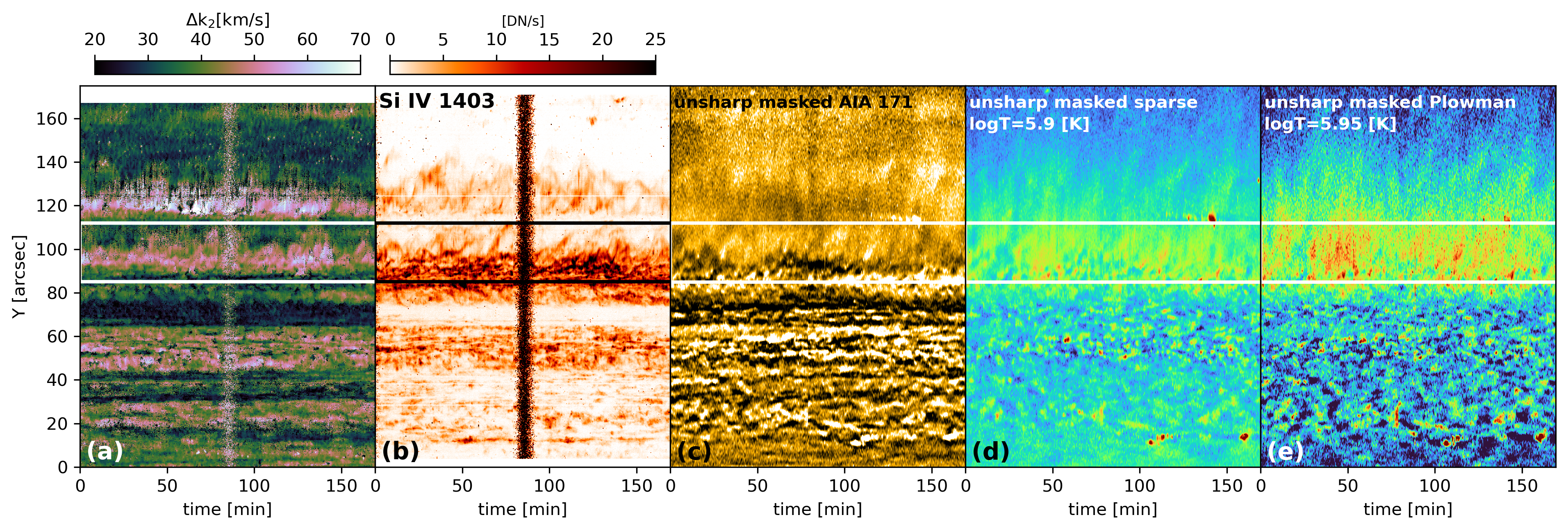}
    \caption{Propagation of spicules in the chromosphere, TR, and lower corona ($\sim$1MK) as observed from the raster in Figure~\ref{fig:gridline2}. \textit{Panel~(a)}: \MgII~k$_{2}$ peak separation. \textit{Panel~(b)}: \SiIV\ 1402.77~\AA\ peak intensity. \textit{Panel~(c):} unsharp masked SDO/AIA~171~\AA\ intensity, \textit{panel~(d)}: unsharped masked EM computed using the method of sparsity, and \textit{panel~(e):} unsharped masked EM computed using Tikhonov regularization. The region bounded by two horizontal solid lines is analyzed further in Figure~\ref{fig_app2:mass_flows_ds2} in appendix~\ref{sec:appendix_supp_figures}.   }
    \label{fig:synth_rast_spectra_ds2}
\end{figure}

Space-time maps generated from the artificial slits marked as $1$, $2$, and $3$ are shown in panels~(e)--(p) of Figs.~\ref{fig:gridline1} and \ref{fig:gridline2}. Slits~1 and 3 in Fig.~\ref{fig:gridline1} lie in the plume region as seen in the 171~\AA\ channel (panel~b) whereas slit 2 along with all the remaining slits in the top row of Fig.~\ref{fig:gridline2} lie outside of plumes. They however lie in close vicinity of the network regions. These space-time maps show abundant bright ridges with predominantly linear shapes, which are consistently visible across all the channels including the \SiIV\ SJI. The linear ridges indicate rapid, outward propagation along the slits. This outward propagation could either be caused by slow-mode magnetoacoustic waves in the low plasma-$\beta$ TR and coronal environment \citep[e.g.,][]{2012A&A...546A..50K} or could be attributed to mass flows as suggested by \cite{2010_bart_outflows}. It is difficult to distinguish between the two possibilities using only imaging data, and high-resolution spectroscopic observations are needed to interpret the picture fully. We refer to Sect.~\ref{sec:discussion_conclusion} for a brief discussion.
% Despite occurring multiple times over the whole duration of the datasets, the nature of these ridges is distinct from the quasi-periodic intensity pulsations that are generally observed in open coronal structures \citep[e.g.][]{2012A&A...546A..50K}. The latter manifests the slow magneto-acoustic waves along the open field lines. 
%
%Moreover, the fact that the ridges in the coronal channel and EM maps have a nearly one-to-one correspondence with the TR \SiIV\ SJI further suggests that these disturbances are not associated with the intensity pulsations but are more aligned with the scenario presented in \cite{2017ApJ...845L..18D,2018ApJ...860..116M}, where they are linked to the TR network jets. 
In addition to the linear trajectories, some instances of parabolic paths traced by the spicules can also be seen in the space-time maps e.g. in Fig.~\ref{fig:gridline2}~(f--h) between 75--85~min, and immediately afterward between 90--105~min indicating the rising and falling phase of spicules. Although more commonly attributed to type-Is, parabolic trajectories lasting between 10 to 15 min are also widely observed in type-II spicules when multi-wavelength observations covering wide range of temperatures (chromospheric to TR/lower coronal) are considered \citep[see, e.g.][]{2014ApJ...792L..15P,2019Sci...366..890S}. The emission (particularly the EM) associated with spicules is found to be substantially enhanced (roughly by a factor of 2 on average) compared with the time intervals where little-to-no spicular activity is observed (e.g. between 20--40~min in Fig.~\ref{fig:gridline1}~e--h) but we notice a gradual drop in their emission as they propagate away from their source. This is also seen in the space-time maps derived from the (synthetic) raster maps (Figs.~\ref{fig:synth_rast_spectra_ds1} and \ref{fig:synth_rast_spectra_ds2}, described below).

The animations associated with the rasters in the left column in Fig.~\ref{fig:synth_rast_spectra_ds1} not only show spicules propagating in the chromospheric \MgII~k$_{3}$ and TR \SiIV~channels (panels~a and b), but also their corresponding propagation in the 171~\AA\ and the two EM maps at $\sim$log~T[K]=$5.9$ (panels c--e). For the sake of brevity, we only show a portion of the raster FOV from dataset~1 indicated by the black bounded region in Fig.~\ref{fig:gridline1}, which has the highest spicule density. To further visualize their propagation we choose two additional (artificial) slits of the same size in the raster maps and show the corresponding space-time plots in panels~(f)--(j). Like the SJI space-time maps, we find linear ridges of enhanced TR and coronal emission associated with spicules that can now also be visualized in the \MgII~k$_{2}$ peak separation space-time maps (panel~f). Due to the relatively low cadence (150~s) of the rasters compared to the SJIs, it is difficult to capture the complete evolution of all spicules many of which last well below 100~s in the chromosphere \citep{2012ApJ...759...18P,2021A&A...647A.147B}. Nonetheless, we see a few examples in both the space-time maps marked with ellipses in the left column of Fig.~\ref{fig:synth_rast_spectra_ds1}. For instance, the enhanced emission ridge between $t=7$--19~min and $t=25$--30~min in the top row, and between $t=45$--50~min and around $t=79$~min in the bottom row shows the propagation in the coronal channel associated with chromospheric and TR counterpart of spicules quite distinctly. Additionally, we find repetitive coronal emission patches (non-ridge-like) between 0--10\arcsec\ in the space-time maps throughout the entire 120~min of observation. These patches are also likely associated with spicules as evidenced by their enhanced \MgII~k$_{2}$ peak separation, but their evolution is not fully captured owing to the low temporal resolution of the rasters. Upper chromospheric velocities derived from the Doppler shift of \MgII~k$_{3}$ and their space-time maps (Fig.~\ref{fig_app1:mass_flows_ds1} in Appendix~A and their animation) show velocities ranging between 20--30~\kms\ (through the shifts in the \MgII~k$_{3}$ core with respect to rest-frame k$_{3}$ wavelength of 2796.352~\AA\ in vacuum\footnote{Source: \url{https://physics.nist.gov/PhysRefData/ASD/lines_form.html}}) at the footpoints of the coronal structure. Being a complex, optically thick spectral line, determining the actual Doppler shifts in \MgII\ associated with spicules is a non-trivial task since opacity effects strongly impacts these profiles \citep{2019A&A...631L...5B}. Nonetheless, the values of Doppler shifts are consistent with recent observations of RBEs and RREs in \MgII, for example, by \cite{2019A&A...631L...5B} and \cite{2023ApJ...946..103H}.

The right column of Fig.~\ref{fig:synth_rast_spectra_ds1} shows two spectral-time (\lambdat, panels~c, and f) diagrams and co-temporal spectrograms (\lambday, panels~b, and e) in \MgII\ and \SiIV\ spectra at the locations marked along the two slits shown in panels~(a) and (d). The slits are chosen based on the location of the space-time analysis discussed in the previous paragraph. Panels~(a), (b), (d), and (e) in the top and bottom right corners of Fig.~\ref{fig:synth_rast_spectra_ds1} are shown at instants indicated in the corresponding \lambdat\ diagrams. The \MgII~k$_{3}$ intensities and \lambday\ spectrograms in Fig.~\ref{fig:synth_rast_spectra_ds1} show the occurrence of many spicules and their associated spectral excursions. The spicule spectra in \MgII\ can readily be identified by the Doppler shift (excursion) of the central absorption k$_{3}$ feature and the corresponding suppression (enhancement) of the respective k$_{2}$ peaks as shown in the inset figures. This is consistent with the analysis presented in \cite{2015ApJ...799L...3R} and \cite{2019A&A...631L...5B}. The \lambdat\ diagrams in Fig.~\ref{fig:synth_rast_spectra_ds1} panel~(c) show the repetitive occurrence of the short-lived asymmetries at the location indicated by a plus sign in panel~(a). The \SiIV\ 1402.77~\angstrom\ spectra are noisier at this exposure (refer to the inset figures), however, comparing panels~(d)--(f) with (a)--(c) and their animation, we find very similar spatio-temporal behavior with blue and redshifts of the line center in tandem with the \MgII~k counterparts. Single Gaussian fits to the \SiIV\ 1402.77~\angstrom\ spectra (see inset panels) reveal Doppler shifts in the range 12--20~\kms.

Analysis of the raster maps derived from dataset~2 shown in Fig.~\ref{fig:synth_rast_spectra_ds2} reveals a consistent scenario where the chromosphere is replete with spicular mass flows that originate in the close vicinity of the network or the footpoint of the coronal plume (between $Y$ = 85--117\arcsec, panel~a). These regions are also naturally accompanied by strong \SiIV\ peak emission shown in panel~(b) owing to the presence of network jets. To accentuate the propagating features, we show an unsharped masked version of the 171~\AA\ and the EM maps (panels~c--e). These panels reveal the ubiquitous signature of linear ridges with a high EM throughout the entire 165~min. Figure~\ref{fig_app2:mass_flows_ds2} in appendix~\ref{sec:appendix_supp_figures} zoom into the above region and shows the presence of strong Doppler shifts in the \MgII~k$_{3}$ associated with areas of enhanced k$_{2}$ peak separation. 
%The impact of viewing angle on these Doppler shifts is even less for this dataset ($\mu$=0.98) compared to dataset~1.

Panels~(a) and (b) in Fig.~\ref{fig:synth_rast_spectra_ds2} show a separate region of spicular activity around $Y$=120\arcsec, which upon examining the \SiIV~SJI in Fig.~\ref{fig:gridline2}~(a) shows the presence of an enhanced spicular activity due to the small patch of network elements. This difference is not obvious in the corona (refer to Fig.~\ref{fig:synth_rast_spectra_ds2}~c--e) due to long, overlying plume structures originating around $Y$=85\arcsec\ seen in the AIA 171 image in Fig.~\ref{fig:gridline2}~(b). However, enhanced emission ridges can be seen in Fig.~\ref{fig:synth_rast_spectra_ds2}~(c)--(d) between $Y$=120--140\arcsec, which are likely their lower coronal counterparts. Interestingly, the region immediately below $Y$=80\arcsec\ is devoid of any spicular ridges in the raster maps, which is further supported by the reduced \MgII~k$_{2}$ peak separation and \SiIV\ peak emission. The \SiIV\ SJI confirms the scenario where spicular features are also absent. The traces of enhanced \MgII~k$_{2}$ peak separation and \SiIV~emission observed between $Y$=20--60\arcsec\ in Fig.~\ref{fig:synth_rast_spectra_ds2} are only due to a (small) portion of the spicules overlapping the IRIS slit and originating around $X$=90\arcsec\ seen in Fig.~\ref{fig:gridline2} panel~(a). Unlike the spicules discussed above, they are not oriented along the slit and hence appear as patches instead of elongated ridges in the space-time raster map in Fig.~\ref{fig:synth_rast_spectra_ds2}.

\section{Discussion and Conclusion} \label{sec:discussion_conclusion}

Ever since the discovery of the more energetic type-II spicules in 2007 \citep{2007PASJ...59S.655D} speculations about their contribution towards heating and mass-loading of the solar corona have been a topic of significant interest and debate \citep[see for example][]{2011Sci...331...55D,2011A&A...532L...1M,2012JGRA..11712102K,2016ApJ...829L..18Bryans,2017ApJ...845L..18D,2019Sci...366..890S,2023ApJ...944..171B}. This paper does not attempt to answer whether spicules play any role in heating the corona as a whole but it presents unique observational evidence that suggests the plasma associated with a subset of spicules can be heated to a million degrees. Based on selection of targets (i.e. QS and CH) and the fact that heating to TR and lower coronal temperatures are involved \citep{2019ARA&A..57..189C}, it is very likely that the spicules investigated in this paper are of the type-II category. The basis of this investigation lies on computing DEM inversions of QS and CH data using two independent algorithms-- one based on $L^{1}$ norm \citep{2015ApJ...807..143C} and the other on $L^{2}$ norm \citep{2020ApJ...905...17P}. The two algorithms differ mainly in minimizing the objective function to obtain the DEMs and therefore serve as an independent way to detect the million-degree emission associated with spicular plasma. The major advantage of this approach is the ability to quantify the amount of emission in a specific temperature bin, which is impossible to infer directly from the AIA images owing to the contamination from cooler TR ions \citep{2011ApJ...743...23M,2018LRSP...15....5D}. To strengthen our claims, we investigate the impact of potential (cooler) TR contamination in the AIA emission associated with network jets using isothermal approximation and filter-ratio diagnostics (cf.~\ref{sec:appendix_lower_T}). Our analysis shows that network jets observed in the AIA passbands can have strong emissions above \(log~T[K]=5.7\) or $\approx$500,000~K, up to at least 1 MK. The representative examples presented in Figs.~\ref{fig_app:network_jet1},  \ref{fig_app:network_jet2}, \ref{fig_app:network_jet3} and \ref{fig_app:filter_ratio} show that even under simplistic assumptions, we can constrain the emission measure in the range \(log~T[K]=[5.5,5.7]\). This further enhances confidence in the results obtained using standard DEM inversions around log~T[K]$\approx$5.9 presented in this paper.

To the best of our knowledge, this is the first time DEM inversions have been applied to study the lower coronal response associated with type-II spicules/network jets in such detail. Recently, \cite{2023_mandal_DF_DEM} applied the method of \cite{2015ApJ...807..143C} to investigate lower coronal response of dynamic fibrils (or type-I spicules) with coordinated IRIS, \SDO\ and Extreme Ultraviolet Imager's Fe~IX~174~\AA\ observations. Though their analysis suggests that some dynamic fibrils may be heated to $>$1~MK, further investigation of the TR contamination and temporal evolution is needed to draw firm conclusions on their exact temperature. Moreover, the dynamic fibrils appear as roundish bright blobs in the 174~\AA\ channel \citep[resembling the ``grains'' in IRIS \SiIV\ passband, ][]{2016_hakon_grains} that have a different morphology than the examples presented in this paper.

Animation of the EM maps, \SiIV\ SJIs and AIA~171~\AA\ images along with the space-time analysis presented in this paper clearly show a spatio-temporal coherence with spicules/network jets observed in the IRIS passband and a strong EM around log~T[K]=5.95 associated with the propagation of the plasma associated with spicules. It is to be noted that the EM obtained from the method of sparsity ($L^{1}$ norm) is on average lower than that obtained from $L^{2}$ norm for a given temperature bin due to the latter's tendency to prefer a flat DEM (in absence of other constraints) compared to minimizing the DEMs (in case of sparsity) as described in \cite{2020ApJ...905...17P}. Nonetheless, both codes provide consistent results where the EM associated with spicular plasma is nearly 2--3 times higher than the respective background emission, which is not observed in the case of dynamic fibrils \citep[e.g.,][]{2023_mandal_DF_DEM}.

Our observations are consistent with a scenario where abundant spicular activity is found in the chromospheric and TR passbands rooted close to the footpoints of the coronal loops \citep{2023ApJ...944..171B}. Moreover, the relationship between spicules setting off PCDs that has been proposed in multiple studies \citep[e.g., in][]{2015_pcds_samanta,2017ApJ...845L..18D,2023_plumes_cho}, is further enhanced. %
Numerical modeling predictions from \cite{2017Sci...356.1269M,2017ApJ...845L..18D} and \cite{2018ApJ...860..116M} suggest that the observed PCDs are likely governed by a complex chain of events involving the generation of spicular flows and associated shock waves, along with heating of plasma through the dissipation of electrical currents and that PCDs are not necessarily caused by magneto-acoustic waves alone. 
Furthermore, the space-time maps (particularly in Figs.~\ref{fig:gridline1} and \ref{fig:gridline2}) show a gradual decrease in the enhancement of the EM associated with PCDs as the disturbances propagate higher up in the corona. This is likely due to the smearing of the density and the temperature of the spicular plasma owing to thermal conduction \citep{2018ApJ...860..116M}.
%\citep[which is in agreement with the predictions of][]{2018ApJ...860..116M}.

\begin{figure}[ht!]
    \centering
    \includegraphics[width=\textwidth,trim={3cm 1cm 3cm 1cm},clip]{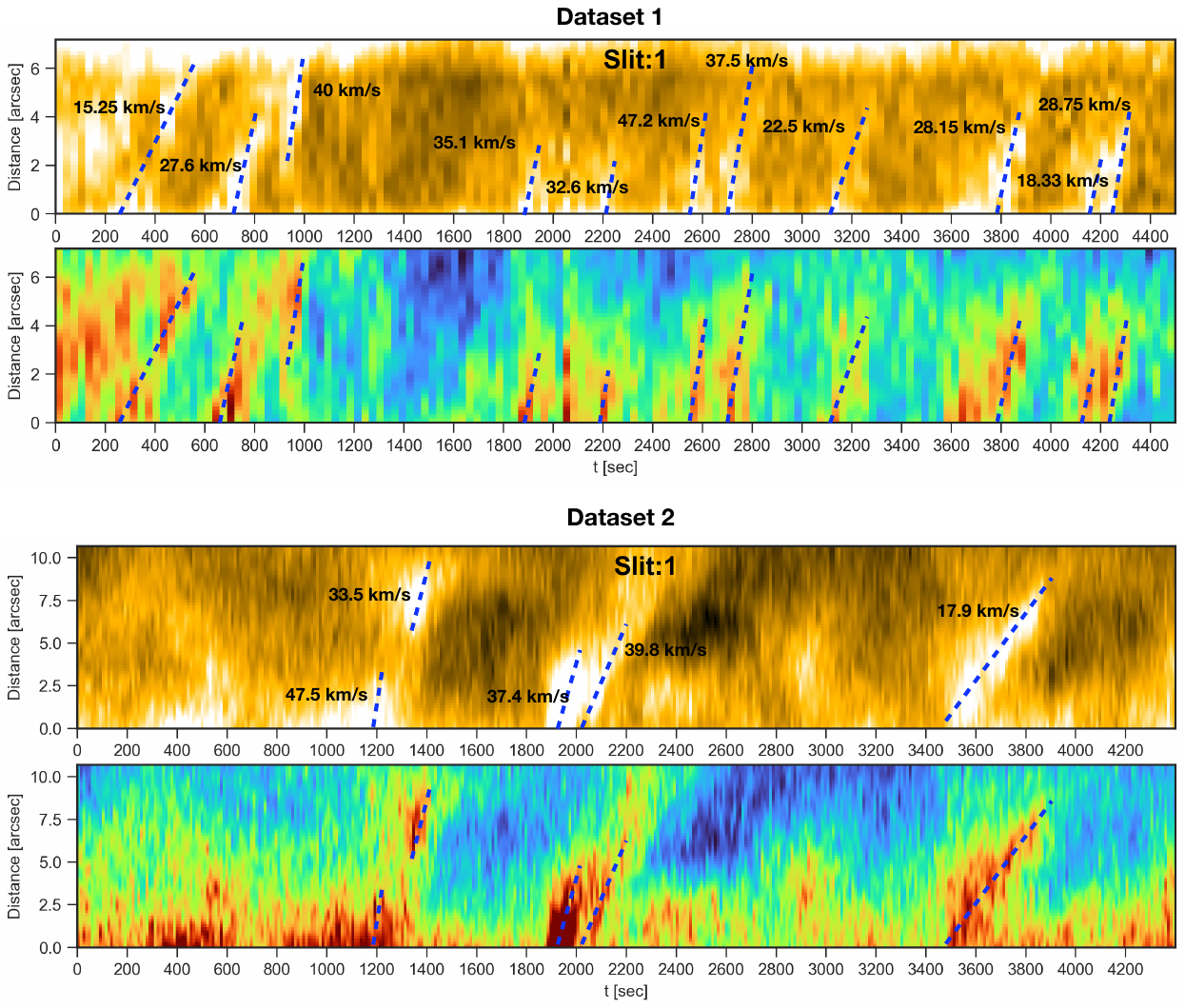}
    \caption{Coronal plane-of-sky velocities corresponding to the x--t maps of slit-1 from datasets~1 (top row) and 2 (bottom row) shown in Figs.~\ref{fig:gridline1} and \ref{fig:gridline2}. The x-axis shows the time in seconds from the beginning of the observation. The dashed blue lines indicate the propagation for which the velocities are indicated. The two panels in each of the two rows correspond to SDO/AIA~171~\AA\ and EM map derived using Tikhonov regularization.}
    \label{fig:wave_vs_flows}
\end{figure}

The animation associated with Fig.~\ref{fig_app1:mass_flows_ds1} in Appendix~\ref{sec:appendix_supp_figures} shows consistent LOS velocities of the order of 20--30~\kms\ in the upper chromosphere along with enhanced \MgII~k$_{2}$ peak separation. The range of the derived Doppler shifts is consistent with the values observed in RBEs and RREs in other chromospheric lines such as \ion{Ca}{ii}~8542~\AA\ and H-$\alpha$ \citep{2013ApJ...769...44S}. The Doppler shifts measured from \SiIV\ are comparable with \cite{2015ApJ...799L...3R} but we note that single Gaussian fitting may not always catch the full complexity of these profiles. Recent studies \citep[e.g.,][]{2018A&A_Kayshap,2022A&A_gorman} suggest that sometimes an extra velocity component $\sim$50--70~\kms\ may be found in the far wings of the \SiIV\ spectral line. In addition, the impact of different viewing angles between the local magnetic field direction or flows, and the LOS may be significant and difficult to determine from a single viewpoint observation. 
% Stronger LOS velocities and peak separation are observed close to the initial stages of the spicule evolution, as evidenced by the space-time maps because the later stages of their evolution are not fully captured owing to the limited temporal cadence of the rasters. This is mitigated to an extent in dataset~2 (Fig.~\ref{fig_app2:mass_flows_ds2}) where the LOS velocities can be tracked over longer distances owing to their higher temporal resolution.
% Nonetheless, the presence of strong chromospheric Doppler shifts at the footpoints of the PCDs combined with the fact that the observations are within proximity to the disk center ($\mu$=0.94 and 0.98 for the two datasets), implies that the disturbances propagating in the lower corona around 1~MK are (likely) associated with mass-flows \citep{2010_bart_outflows} and are not caused due to the rapidly ($>$100~\kms) propagating heating fronts \citep{2017_bart_heating_fronts}. 
Interestingly, from Fig.~\ref{fig:wave_vs_flows} we notice that the apparent velocities of the coronal signal projected into the plane-of-sky are primarily between 15--45~\kms, and that the Doppler shifts in k$_{3}$ of \MgII\ is similar. We note that the actual Doppler shifts of the chromospheric plasma may be higher than those derived from the k$_{3}$ core spectral feature since that is part of a complex profile in an optically thick line. 

If mass flows indeed cause the plane-of-sky motions, they would be compatible with a scenario in which real flows of order 40~\kms\ are observed with a viewing angle of order 45 degrees (between the magnetic field vector and the LOS), leading to Doppler shifts in the upper chromosphere of 20--30~\kms\ (as seen in \MgII~k$_{3}$), and projected velocities in the plane-of-sky in coronal images of 20--30~\kms. Alternatively, if the apparent coronal motions are caused, by, e.g., a sound wave of thermal conduction front \citep[e.g.,][]{2017_bart_heating_fronts}, the viewing angle would be 60 degrees or higher since field-aligned speeds of waves or conduction fronts are expected to be of order 100~\kms\ or more. 

The \lambdat\ and \lambday\ slices of the \SiIV\ spectra shown in the right column of Fig.~\ref{fig:synth_rast_spectra_ds1} along with its animation show similar spatio-temporal behavior with blue (red) shifted excursions in tandem with the \MgII~k line. This behavior reaffirms the findings of \cite{2015ApJ...799L...3R} and suggests that the network jets are the TR counterparts of chromospheric RBEs and RREs.

%The skepticism over the role of spicules in energizing the solar corona is still widespread in the community. The concerns are valid to some extent because spicules are an enigma, and coordinated, high-resolution observations covering multiple layers of the solar atmosphere (at high cadence) are needed to fully understand their impact on outer atmospheric heating. This is not an easy task because the spatio-temporal properties of spicules are often at the limit of many current instrumentation capabilities. 

The results presented in this paper highlight the multithermal nature of type-II spicules \citep[similar to][to name a few]{2011Sci...331...55D,2014ApJ...792L..15P,2021_chintzoglou_spicule,2021A&A...654A..51B} along with their complex spatio-temporal evolution where a significant fraction of the plasma associated with spicules is heated to a million degrees with a lower temperature threshold of about 500,000~K. This conclusion is based on a comprehensive DEM analysis of two (quiet Sun and CH) datasets using two independent algorithms. This is the first time multiple DEM-based approaches have been applied to type-II spicules to study their emission in the lower corona in such detail. Our results are compatible with predictions from previously reported advanced numerical models \citep{2017Sci...356.1269M,2018ApJ...860..116M,2020ApJ...889...95M} that show heating to TR and coronal temperatures associated with spicules. Moreover, the computed synthetic TR and coronal images, e.g. in \cite{2018ApJ...860..116M}, show remarkable similarities to the brightenings seen in TR and coronal passband observed in the current study. The generation of the simulated type-II spicules in the above papers drive  Alfv\'enic waves and electric currents that travel along the magnetic field. The dissipation of such waves and/or currents lead to heating of the associated plasma to TR and coronal temperatures, and the whole process leads to synthetic PCDs similar to our observations. This paper is therefore a step ahead of previous studies (owing to the DEM-based approaches and an independent determination of the contribution from cooler TR ions), and the results do not exclude the possibility that spicules do indeed play a role in energizing the lower solar corona, which remains a widely debated topic in the community. To fully resolve this issue will require coordinated, high-resolution observations covering multiple layers of the solar atmosphere (at a high cadence). This is not an easy task because the spatio-temporal properties of spicules are often at the limit of many current instrumentation capabilities. High-resolution coronal spectroscopic observations from the upcoming NASA's MUlti-Slit solar Explorer \citep[MUSE,][]{2020ApJ...888....3D,2022_physics_muse_bdp} mission, Solar-C EUVST, in coordination with IRIS and ground-based data such as DKIST and SST would be the obvious next step in understanding if any heating along coronal loops is linked to spicular injection at its base. MUSE's comprehensive spectroscopic coverage, encompassing the entire length of coronal loops and their footpoints, would allow for reconstructing their complete thermodynamic history. This includes the preceding phase of spicule activity, the subsequent dissipation of currents or waves within the loops, and the final cooling stages.
We look forward to such developments.

\begin{acknowledgments}
IRIS is a NASA small explorer mission developed and operated by LMSAL with mission operations executed at NASA Ames Research Center and major contributions to downlink communications funded by ESA and the Norwegian Space Centre. S.B. gratefully acknowledges support from NASA's SDO/AIA contract (NNG04EA00C) and S.B. and B.D.P. acknowledge support from the IRIS contract (NNG09FA40C) to LMSAL. J.J. acknowledges funding support from the SERB-CRG grant (CRG/2023/007464) provided by the Anusandhan National Research Foundation, India. PT is funded for this work by contract 8100002705 (IRIS) to the Smithsonian Astrophysical Observatory. We acknowledge fruitful discussions with Marc De Rosa while preparing the manuscript.
\end{acknowledgments}

\vspace{5mm}
% \facilities{HST(STIS), Swift(XRT and UVOT), AAVSO, CTIO:1.3m,


\software{SunPy \citep{Barnes2020},  
          SciPy \citep{2020SciPy-NMeth}, Matplotlib \citep{Hunter:2007}, Numpy \citep{harris2020array},
          SSWIDL
          % Source Extractor \citep{1996A&AS..117..393B}
          }



\appendix
\section{Determining the lower temperature cutoff in network jets}
\label{sec:appendix_lower_T}

\begin{figure}[h!]
    \centering
    % \plotone{Figures/AIA_IRIS_RFs.png}
    \includegraphics[width=0.5\linewidth]{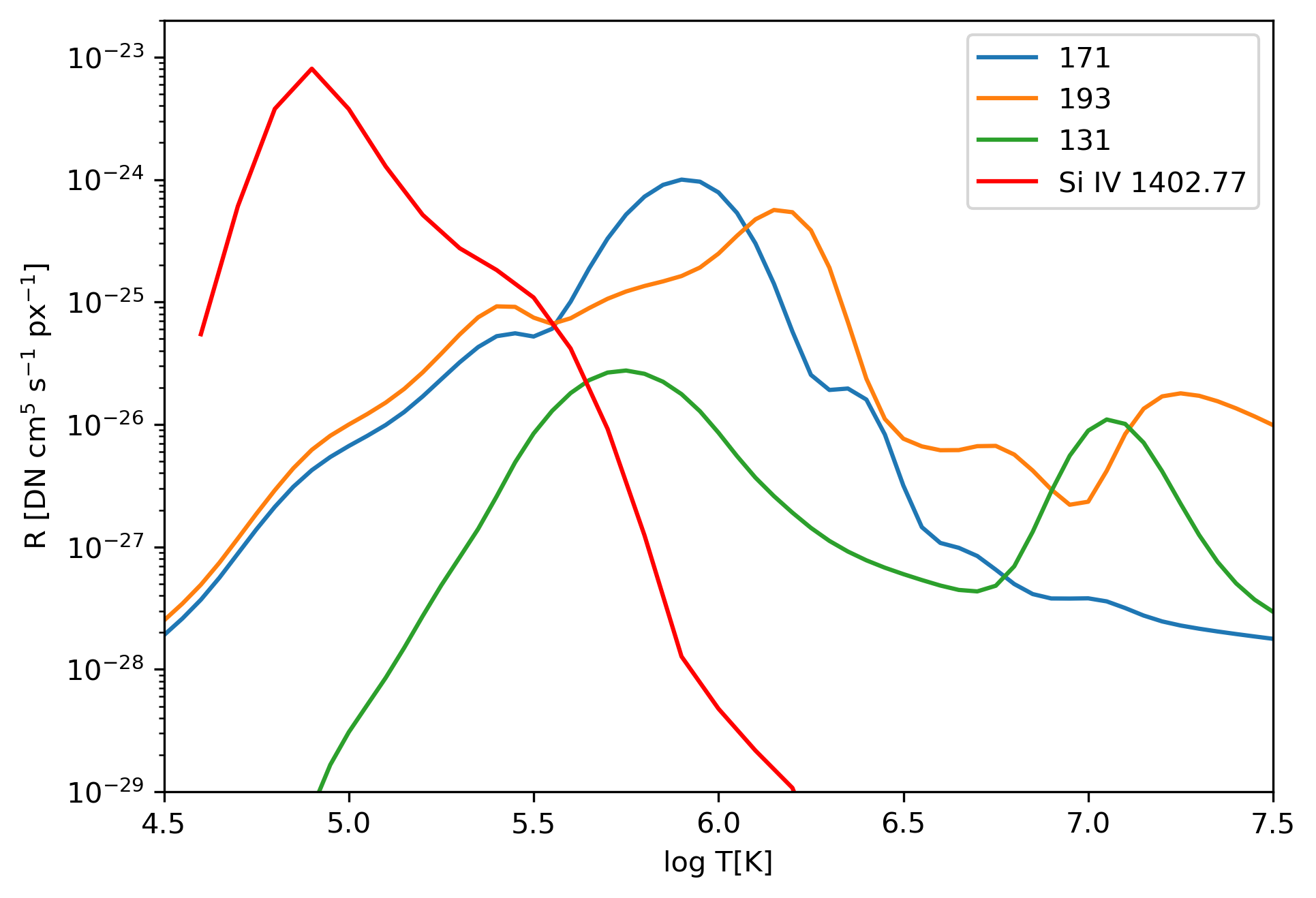}
    \caption{Response functions for \SiIV\ and SDO/AIA 131, 171 and 193~\AA\ EUV channels. The response functions for the (selected) AIA channels are calculated using the routines available in SSWIDL and the \SiIV\ line is calculated using the method described in the appendix. }
    \label{fig_app_RFs:IRIS_AIA_RFs}
\end{figure}

In this appendix, we discuss two approaches that we used to constrain and estimate the contribution of the lower transition region to the DEM analysis, given its potential contributions to the AIA passbands. The ambiguity in DEM results from AIA is caused by the known TR contamination of the AIA passbands. We used IRIS observations and applied two methods to help constrain this contribution: an isothermal approximation of the plasma as well as filter-ratio diagnostics. In both cases, we compute the EM from the two methods (EM$_{\mathrm{derived}}$), and compute the predicted intensity (counts) associated with network jets as would be observed in IRIS \SiIV\ SJI assuming it is dominated by \SiIV~1402.77~\AA. In principle, we use
\begin{equation}
    I_{\mathrm{pred}}=EM_{\mathrm{derived}}\times R
    \label{equation:DEM}
\end{equation}

where $R=R_{\mathrm{Si~IV}}$ is the response function of the \SiIV~1402.77~\AA\ line computed using \verb|isothermal.pro| routine available in CHIANTI v10.1 \citep{2023_CHIANTI_Dere} in SSWIDL. We assumed coronal abundances \citep[sun\_coronal\_2021\_chianti.abund from][]{2021_abundance_Asplund}, an electron number density of 10$^{10}$~cm$^{-3}$ that is typical in \SiIV\ around log~T[K]=4.9 under non-flaring conditions \citep[e.g.][]{2018_Young}. To convert the synthetic \SiIV\ spectra from physical [Photon~cm$^{-2}$s$^{-1}$sr$^{-1}$px$^{-1}$] units to [DN~s$^{-1}$~px$^{-1}$], we use the effective areas and photon to DN conversion values of IRIS SJI in \SiIV\ from \verb|iris_get_response.pro|. The response functions ($R$) of the \SiIV\ line and selected SDO/AIA channels are shown in Fig.~\ref{fig_app_RFs:IRIS_AIA_RFs}.

% \subsection{Predicted Si~IV counts assuming isothermal approximation approach}
\paragraph{\textbf{Isothermal Approximation}} Assuming that the emitting plasma is isothermal, equation~\ref{equation:DEM} can be used to obtain the total source EM by dividing the observed intensity by the response function at a given temperature. The following steps determine the lower temperature cutoff associated with spicular plasma emitted in log~T[K]=5.9. We:

\begin{enumerate}
    \item Calculate the $EM$ at \(log~T[K]=5.9\) (say \(EM_{5.9}\)) by dividing the observed intensity (in a network jet) by the respective AIA filter response functions ($R_{5.9}$) at this temperature.
    
    \item Consider a hypothetical scenario where the emission observed above has been incorrectly assigned to $EM$ at \(log~T[K]=5.9\) but is instead due to contribution from a lower temperature (e.g., \(log~T[K]=4.9\)). To satisfy this requirement, \(EM_{4.9}=EM_{5.9} \times \frac{R_{5.9}}{R_{4.9}}\), where $R_{4.9}$ is AIA filter response at \(log~T[K]=4.9\).
    
    \item Use equation~\ref{equation:DEM} to compute predicted \SiIV~1402.77~\AA\ intensity at \(log~T[K]=4.9\) using the response function shown in Fig.~\ref{fig_app_RFs:IRIS_AIA_RFs} and the derived $EM_{4.9}$ to compare the result with observed intensity in IRIS~\SiIV~1400~\AA\ SJI.
    
    \item If the predicted intensity (at \(log~T[K]=4.9\)) is comparable to the observed value (with IRIS), it is likely that the observed intensity in the AIA channels is due to the emission from the cooler TR component observed by IRIS. Otherwise, if the predicted intensity is much larger than the observations, the cooler TR contamination in AIA is unlikely.
    
\end{enumerate}

\begin{figure}
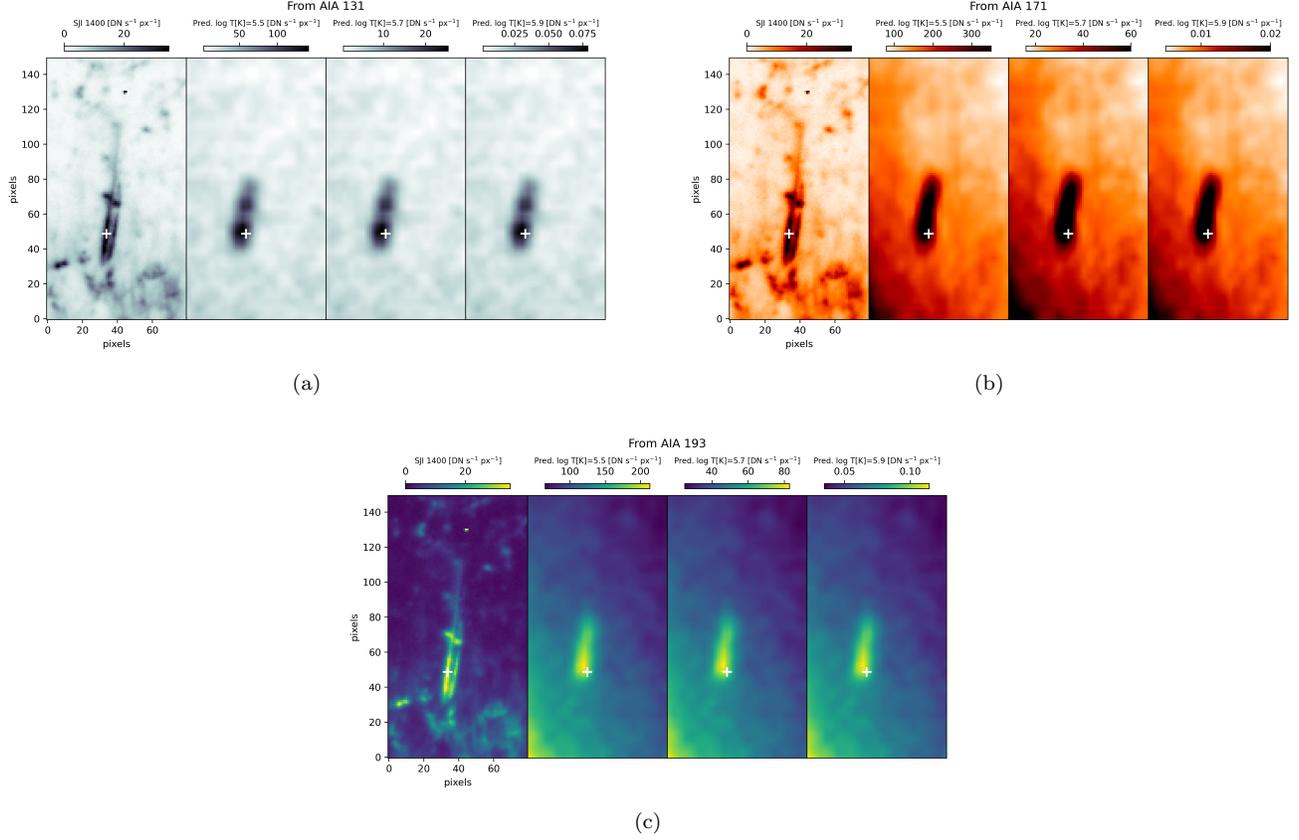

    \gridline{\fig{predicted/network_jet2_from_131.png}{0.45\textwidth}{(a)}
    \fig{predicted/network_jet2_from_171.png}{0.45\textwidth}{(b)}}
    \gridline{\fig{predicted/network_jet2_from_193.png}{0.45\textwidth}{(c)}}
    \caption{A representative example of a network jet observed in coordinated IRIS and SDO/AIA dataset~1 to illustrate the (cooler) TR contamination in the AIA passbands. Predicted counts in different temperature bins obtained under isothermal approximation are shown for AIA~131~\AA\ in panel~(a), AIA~171~\AA\ in panel~(b), and AIA~193~\AA\ in panel~(c). The leftmost image in each of the panels shows the observed IRIS \SiIV\ 1400 counts for comparison and the ``plus'' marker serves as a guide for locating the same pixel across the different images. Note the different color bars in each of the images, which were intentionally chosen to optimize the visualization of the network jet. The size of each pixel is 0\farcs{167} in the horizontal and vertical direction.}
    \label{fig_app:network_jet1}
\end{figure}

\begin{figure}
    \gridline{\fig{predicted/network_jet1_from_131.png}{0.45\textwidth}{(a)}
    \fig{predicted/network_jet1_from_171.png}{0.45\textwidth}{(b)}}
    \gridline{\fig{predicted/network_jet1_from_193.png}{0.45\textwidth}{(c)}}
    \caption{Same as \ref{fig_app:network_jet1} but for a different network jet observed in dataset~1}
    \label{fig_app:network_jet2}
\end{figure}

\begin{figure}
    \gridline{\fig{predicted/network_jet3_from_131.png}{0.45\textwidth}{(a)}
    \fig{predicted/network_jet3_from_171.png}{0.45\textwidth}{(b)}}
    \gridline{\fig{predicted/network_jet3_from_193.png}{0.45\textwidth}{(c)}}
    \caption{Same as \ref{fig_app:network_jet1} but for a different network jet observed in dataset~2.}
    \label{fig_app:network_jet3}
\end{figure}

The above steps are repeated for a range of temperature values between \(log~T[K]=[4.9,5.9]\) in intervals of 0.1, for the three AIA channels (131,171, and 193) over the whole FOV. This analysis did not include AIA 211, 335, and 94~\AA\ because of their relatively low sensitivity below \(log~T[K]=5.7\). We show three illustrative examples of network jets in Figs.~\ref{fig_app:network_jet1}, \ref{fig_app:network_jet2}, and \ref{fig_app:network_jet3}, where we follow the approach outlined above and show a comparison of the observed IRIS \SiIV\ SJI and predicted intensities in three temperature bins i.e. \(log~T[K]=5.6\), $5.7$, and $5.9$. The examples clearly indicate that the predicted range of intensities at \(log~T[K]=5.7\) best resemble the IRIS SJI~1400~\AA\ observations, while at temperature bins lower and higher than \(log~T[K]=5.7\) a clear mismatch with observations is seen. Moreover, the AIA signal does not extend over the same spatial range as the IRIS observations suggesting that the predicted intensities are not a simple print-through of the IRIS passband. In other words, the emission in the AIA passbands cannot be only due to the cooler TR contaminants. Therefore, this analysis strongly suggests that the lower temperature cutoff (or the relatively cool TR contamination) in the network jets observed in the AIA channels cannot be below \(log~T[K]=5.7\) or $\approx$500,000~K.

\paragraph{\textbf{Filter-ratio diagnostic}} The filter-ratio (of the different AIA temperature response functions) is a popular diagnostic that has been extensively used \citep[see e.g.,][for a detailed discussion]{2011_Narukage_filter_ratio} to provide a fast and an approximate way to determine the coronal temperatures using the  
observed intensity ratio in the corresponding filters. Here, we use the filter-ratio in the AIA 171 and 131 passbands within \(log~T[K]=[5.5,6]\) to determine the temperature. The choice of this range is based on the (approximate) linear dependence of the filter-ratio with temperature, which allows for a unique determination of the temperature. Once the temperature is determined, the total EM can be determined by dividing the observed intensity in 171 (or 131) by the corresponding response ($R$) at that temperature. Finally, the total EM is folded with the $R$ of \SiIV~1402.77~\AA\ using equation~\ref{equation:DEM} to obtain the predicted intensities.

\begin{figure}
    \centering
    \includegraphics[width=\textwidth]{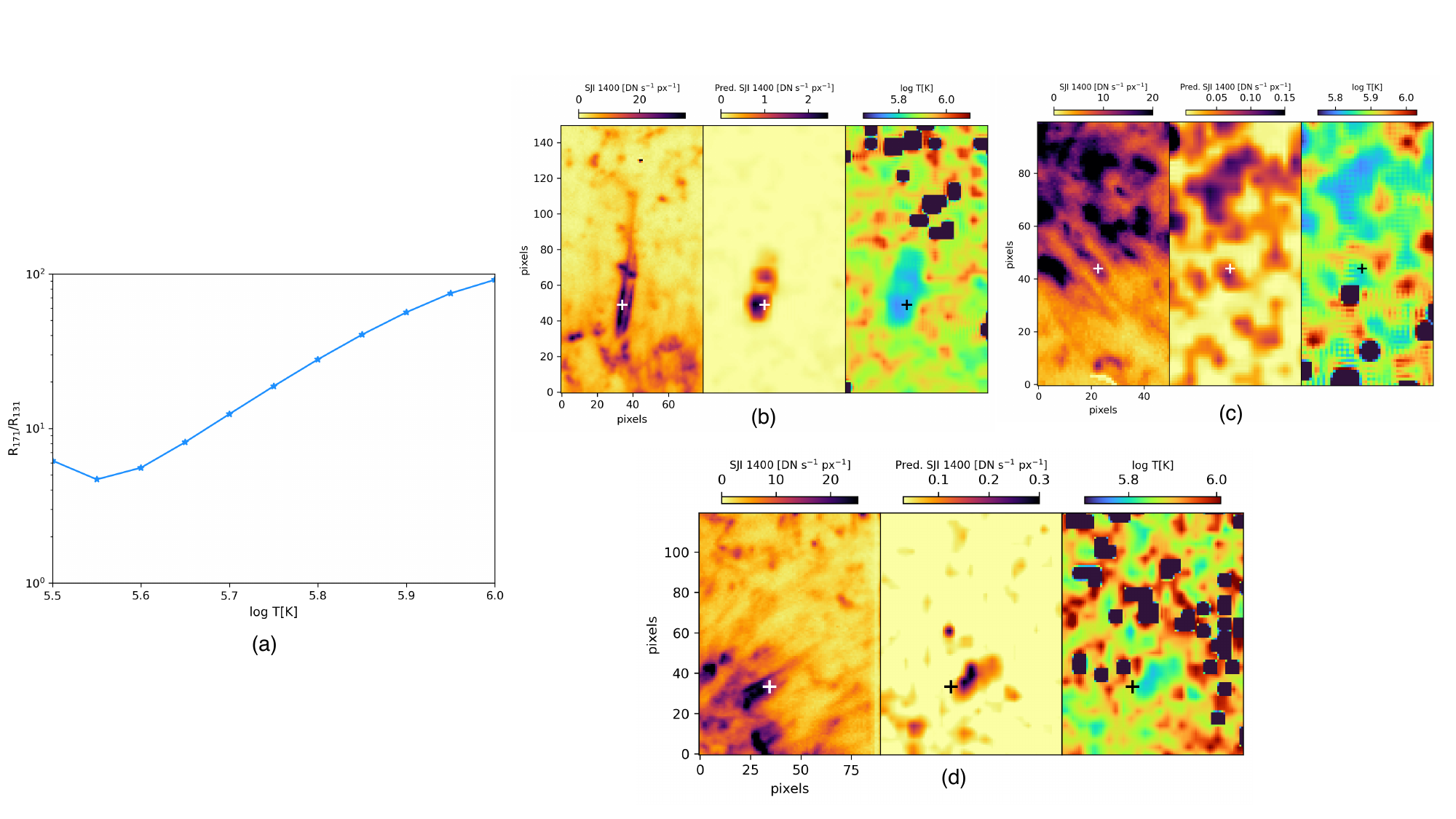}
    \caption{Temperature diagnostics with the filter-ratio method. Panel(a): Ratio of the temperature response of the SDO/AIA 171 and 131 filters between \(log~T[K]=[5.5,6.0]\). Panel~(b), left to right: observed IRIS \SiIV~1400 SJI intensities for network jet 1, predicted intensities using the total EM, and temperature map derived from the filter-ratio technique. Panels~(c) and (d) are the same as panel~(b) but for network jets~2 and 3. The FOV and the pixel size are the same as Figs.~\ref{fig_app:network_jet1}, \ref{fig_app:network_jet2}, and \ref{fig_app:network_jet3}. }
    \label{fig_app:filter_ratio}
\end{figure}

Figure~\ref{fig_app:filter_ratio} panel~(a) shows the AIA 171 and 131 passbands' filter-ratio and temperature dependence. Between \(log~T[K]=[5.55,6.0]\) the ratio is approximately linear and monotonically increasing which allows us to uniquely determine the temperature of the three network jets shown in panels~(b)--(d). The morphology of the network jets appears similar in the maps of predicted intensities and temperature. They allow us to conclude that the minimum temperature associated with the three network jets is around \(log~T[K]=5.7\), which corresponds with the minimum temperature values derived from the isothermal approximation approach in the previous section. Interestingly, we also find a variation of the derived temperature along the length of the network jets~1 and 3 where temperatures around \(log~T[K]=5.9\) are also seen.

\section{Supplementary figures}
\label{sec:appendix_supp_figures}

\begin{figure*}[htpb!]
    \plottwo{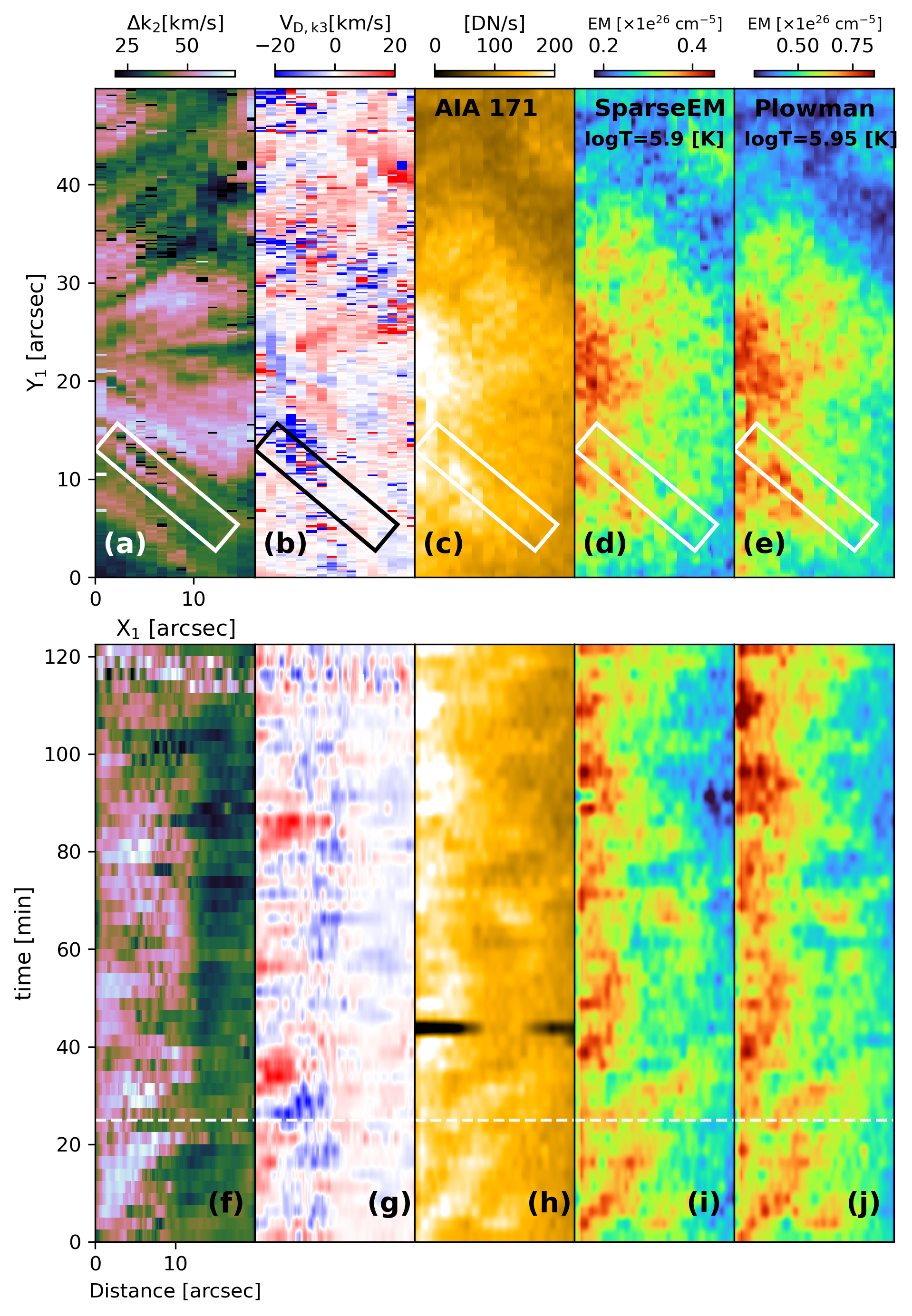}{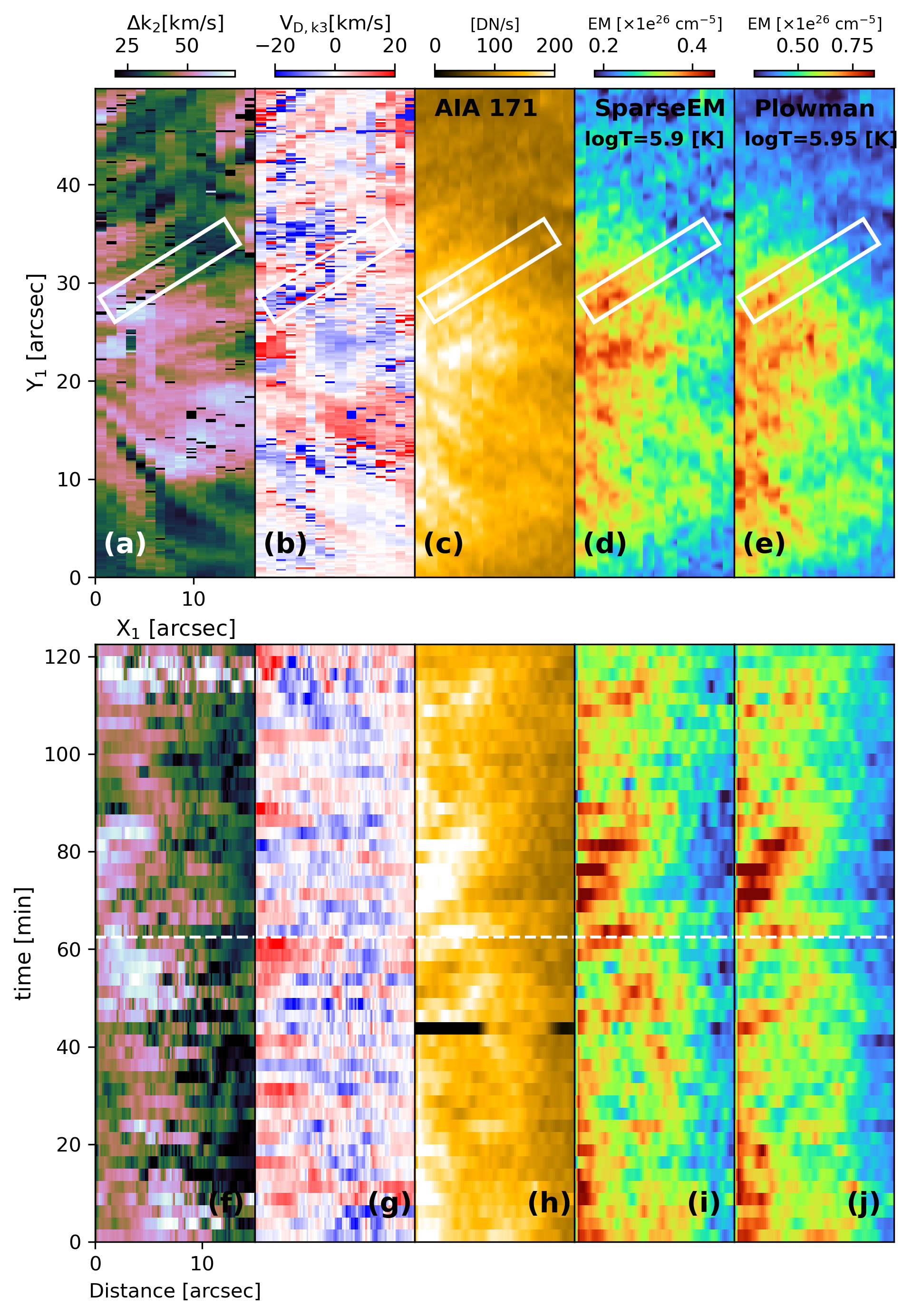}
    \caption{Connecting chromospheric Doppler shifts and emission in the lower corona for the same FOV as in Figure~\ref{fig:synth_rast_spectra_ds1}. \emph{Left column,} (a): \MgII~k$_{2}$ peak separation, (b): Doppler shift of the line core (k$_{3}$), \emph{(c)--(e)}: synthetic SDO/AIA~171, and EM rasters corresponding to the IRIS rasters. The rectangular boxes in \textit{(a)--(e)} show the FOV from which the space-time maps shown in \textit{panels~(f)--(j)} are derived. \emph{Right column:} Same as the left column except for the bottom row \emph{panels~(f)--(j)}, which shows the space-time maps corresponding to the other rectangular box in the top row. An animation showing the temporal evolution of the two columns is available \href{https://www.dropbox.com/scl/fo/5azkx0vky8p0x47dk1yzu/AB_3qOYfAvk8IG3Xy6iJYTI?rlkey=zxm2j4u4mx2ucitvv4wraylpe&st=msqystju&dl=0}{here}.
    \label{fig_app1:mass_flows_ds1}}
\end{figure*}

\begin{figure}
    \plotone{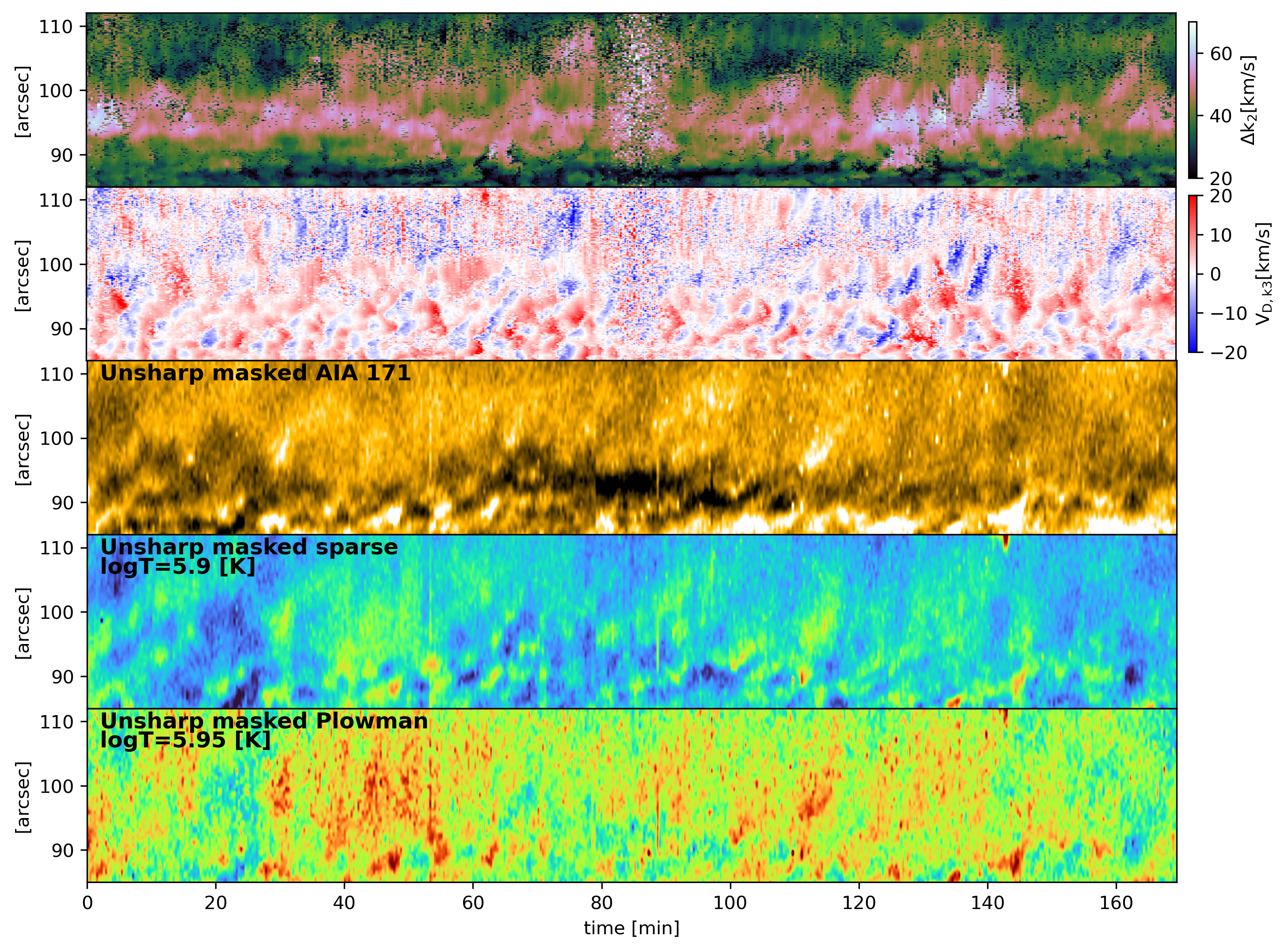}
    \caption{A zoom-in to the chromospheric Doppler shifts and lower coronal emission associated with spicular plasma for the bounded region indicated in Fig.~\ref{fig:synth_rast_spectra_ds2}. Rows from top to bottom show \MgII~k$_{2}$ peak separation map, Doppler shift of the line core (k$_{3}$), unsharped masked version of SDO/AIA~171, EMs computed using the method of sparsity, and Tikhonov regularization. }
    \label{fig_app2:mass_flows_ds2}
\end{figure}

\bibliography{sample631}{}
\bibliographystyle{aasjournal}

\end{document}